\documentstyle[epsfig,natbib2,natbibmnfix]{mn}

\newcommand{\be}{\begin{equation}}
\newcommand{\ee}{\end{equation}}
\newcommand{\bea}{\begin{eqnarray}}
\newcommand{\eea}{\end{eqnarray}}
\newcommand{\bc}{\begin{center}}
\newcommand{\ec}{\end{center}}

\renewcommand{\vec}[1]{ {\bmath #1} }

\def\gsim{ \lower .75ex \hbox{$\sim$} \llap{\raise .27ex \hbox{$>$}} }
\def\lsim{ \lower .75ex \hbox{$\sim$} \llap{\raise .27ex \hbox{$<$}} }

\renewcommand{\thefootnote}{\fnsymbol{footnote}}

\title{Physical viscosity in smoothed particle hydrodynamics simulations of
  galaxy clusters}

\author[Sijacki \& Springel]
       {\parbox{18cm}{Debora~Sijacki\footnotemark[1] and
       Volker~Springel}\vspace{0.3cm}\\ 
       Max-Planck-Institut f\"{u}r Astrophysik,
       Karl-Schwarzschild-Stra\ss{}e 1, 85740 Garching bei M\"{u}nchen, Germany}

\begin{document}

\maketitle
\begin{abstract}
  Most hydrodynamical simulations of galaxy cluster formation carried
  out to date have tried to model the cosmic gas as an ideal, inviscid
  fluid, where only a small amount of (unwanted) numerical viscosity
  is present, arising from practical limitations of the numerical
  method employed, and with a strength that depends on numerical
  resolution.  However, the physical viscosity of the gas in hot
  galaxy clusters may in fact not be negligible, suggesting that a
  self-consistent treatment that accounts for the internal gas
  friction would be more appropriate. To allow such simulations using
  the smoothed particle hydrodynamics (SPH) method, we derive a novel
  SPH formulation of the Navier-Stokes and general heat transfer
  equations and implement them in the {\small GADGET-2} code.  We
  include both shear and bulk viscosity stress tensors, as well as
  saturation criteria that limit viscous stress transport where
  appropriate. Our scheme integrates consistently into the entropy and
  energy conserving formulation of SPH employed by the code.  Using a
  number of simple hydrodynamical test problems, e.g.~the flow of a
  viscous fluid through a pipe, we demonstrate the validity of our
  implementation. Adopting Braginskii parameterization for the shear
  viscosity of hot gaseous plasmas, we then study the influence of
  viscosity on the interplay between  AGN--inflated bubbles and the
  surrounding intracluster  medium (ICM). We find that certain bubble
  properties like morphology, maximum clustercentric radius reached,
  or survival time depend quite  sensitively on the assumed level of
  viscosity. Interestingly, the sound waves launched into the ICM by
  the bubble injection are damped by physical viscosity, establishing
  a non-local heating process. However, we find that the associated
  heating is rather weak due to the overall small energy content of
  the sound waves.  Finally, we carry out cosmological simulations of
  galaxy cluster formation with a viscous intracluster medium.  We
  find that the presence of physical viscosity induces new modes of
  entropy generation, including a significant production of entropy in
  filamentary regions perpendicular to the direction of the clusters
  encounter. Viscosity also modifies the dynamics of mergers and the
  motion of substructures through the cluster
  atmosphere. Substructures are generally more efficiently stripped of
  their gas, leading to prominent long gaseous tails behind infalling
  massive halos.
\end{abstract}

\section{Introduction}

\renewcommand{\thefootnote}{\fnsymbol{footnote}}
\footnotetext[1]{E-mail: deboras@mpa-garching.mpg.de }

Studies of the intracluster medium (ICM) provide unique information about the
complex interplay of the physical processes that determine the fate of baryons
in galaxy groups and clusters. In recent years, remarkable observational
progress has in fact unveiled a completely revised picture of the intracluster
medium (ICM) where a plethora of non-gravitational physical processes are
responsible for key observational phenomena. Indeed, the list of recent
discoveries in the field of ICM physics is quite long, and includes cold
fronts, long X-ray tails in the wake of late type galaxies passing through the
hot cluster environment, or the presence of radio halos and ghosts associated
with past AGN activity. Theoretical studies of these phenomena
increasingly rely on direct numerical simulations, which are in principle
capable of accurately computing the non-linear interplay of all these
processes and their consequences for the thermodynamics of the ICM. A
prerequisite is that the simulations are capable of representing all the
physics relevant for the system, which represents a significant ongoing
challenge.

A particularly important question in cluster physics concerns the
observed absence of strong cooling flows onto the massive elliptical
galaxies at the centres of the potential wells of groups and clusters
of galaxies \citep[e.g.][]{Peterson2001, Peterson2003, Tamura2001,
Balogh2001, Edge2001, Edge2002, Boehringer2002}.  There is now growing
observational evidence for the relevance of AGN heating in these
objects \citep[e.g.][]{McNamara2000, Sanders2002, Mazzotta2002,
McNamara2005,Fabian2006}, supporting the widespread theoretical notion
that the central AGN is providing enough energy to offset the
radiative cooling losses.  However, it is still not understood in
detail how this energy is coupled into the ICM. X-ray observations
with the XMM-Newton and Chandra telescopes \citep[e.g.][]{Blanton2001,
Birzan2004, Nulsen2005} have revealed that in many cooling-flow
clusters there are so called X-ray cavities which interact with the
surrounding intracluster gas. It is believed that these bubbles are
inflated by the powerful AGN jets that are generated by an accreting
central black hole. The expanding bubbles heat the ICM by mechanical
work, and by the buoyant uplifting of cool gas from the central
regions and subsequent mixing with the hotter atmosphere at larger
radii. At the same time, the bubbles trigger sound waves that travel
through the cluster, and they may excite global oscillations modes of
the ICM in the cluster potential. It has been suggested that viscous
damping of these sound waves may provide an important non-local
heating source for the ICM \citep[e.g.][]{Fabian2003}. A significant
cluster viscosity may in principle exist, but its strength should
depend critically on the magnetic field strength and the field
topology.

Radio observations \citep[e.g.][]{Owen2000, Clarke2005, Dunn2005} have clearly
shown that the X-ray cavities are filled with relativistic gas, and probably
have inherited some of the magnetic fields transported by the AGN jet. While
the structure of the magnetic fields filling the bubbles is not well known
yet, it has by now been firmly established that galaxy clusters are permeated
by magnetic fields \citep[for reviews see][]{Carilli2002, Govoni2004}. Faraday
rotation measurements \citep{Clarke2001, Clarke2004, Eilek2002, Vogt2003,
  Vogt2005} have found that the magnetic fields in clusters appear to be
random, with an rms strength of order of $1-10\, \mu{\rm G}$, and with a
coherence length of $1-20\, {\rm kpc}$.  Assuming a fully ionized plasma
\citep{Spitzer}, this implies a very high magnetic Prandtl number for the ICM
of order of $\sim 10^{29}$, suggesting that magneto-hydrodynamic (MHD)
turbulence is probably relevant. On the other hand, the inferred typical
Reynolds numbers for the intracluster gas are quite small, $\lsim 100$,
indicating that the gas viscosity might be quite important.  Moreover, the
coherence length-scale of the magnetic fields is comparable to the ion mean
free path, as well as to the typical size of galaxies and of AGN--driven
bubbles that could drive turbulence in the ICM.  In fact, there have been some
observational studies that found evidence for the presence of turbulence in
clusters \citep{Schuecker2004,Rebusco2005}, suggesting that the injection
scale of the turbulence would be of order $\sim 10-100\, {\rm kpc}$.

All these observational pieces of information do not yet combine to a
definitive picture of the magnetic and viscous properties of the ICM,
and the theoretical understanding is also not yet mature \citep[for a
recent review see][]{Schekochihin2005}. It is clear, however, that an
approximation of the ICM as an ideal, inviscid gas -- as usually made
in most hydrodynamical simulations of galaxy cluster formation -- may
be a poor approximation for hot clusters with non-negligible (and
perhaps chaotically tangled) magnetic fields. It is therefore the aim
of this work to explore the potential imprints of gas viscosity on
galaxy cluster properties in a fully self-consistent way, using
cosmological simulations of cluster growth from $\Lambda$CDM initial
conditions. As a prerequisite for such simulations, we develop a
numerical scheme capable of accurately solving the Navier-Stokes
equations in SPH, which we use instead of the commonly employed much
simpler Euler equation.  We will assume a simple parameterization of
the shear and bulk viscosity tensors, without explicitly dealing with
the MHD equations. For this purpose, we adopt Braginskii
parameterization \citep{Braginskii1958, Braginskii1965} of the
shear viscosity, together with a phenomenological suppression factor
to mimic the influence of the magnetic fields. The bulk viscosity
coefficient is kept constant if included.

In order to test our new hydrodynamical scheme, we apply it to a number of
simple test problems with known analytic solutions. These tests yield robust
results in agreement with the expectations.  We then carry out simulations of
galaxy clusters where we include different physics, with or without physical
viscosity. These simulations include models with non-radiative hydrodynamics,
and models with radiative cooling, star formation and supernovae (SNe)
feedback, allowing us to obtain an overview about the interplay of  gas
dynamics and viscous dissipation in different environments, and the
consequences viscosity has for the structure of clusters.  We also try to
identify potential observational signatures for internal friction processes.

In an additional set of simulations, we analyze the impact of gas viscosity on
the AGN--driven bubble heating process. Previous analytic work
\citep[e.g.][]{Kaiser2005} and numerical Eulerian simulations
\citep[e.g.][]{Ruszkowski2004,Reynolds2005} have suggested that internal
friction has a significant impact on bubble properties, stabilizing them
against hydrodynamical instabilities that would otherwise readily disrupt
them. We explore this issue in some detail with our simulation methodology, in
particular studying bubble morphologies, maximum clustercentric distance, and
survival times, as a function of the assumed level of shear viscosity.  In
this context we also examine the total energy in the sound waves triggered by
the bubbles. We find that this is rather small, something that could in part be
caused by deficiencies in our models, as we discuss later on.

The outline of this paper is as follows. In Section~\ref{Theoretical}, we
review the fundamental physical laws of viscous fluids, focusing in particular
on astrophysical plasmas and a discussion of the role of magnetic fields. The
detailed description of our numerical implementation of physical viscosity in
SPH is given in Section~\ref{Numerical}, while we illustrate the validity of
our numerical scheme with a number of basic hydrodynamical test problems in
Section~\ref{Illustrative}. In Section~\ref{AGN--driven}, we analyze the
heating effects caused by AGN--driven bubbles rising through viscous
intracluster gas, while in Section~\ref{Cosmological}, we discuss internal
friction during merging episodes and its impact on substructure motion in
cosmological simulations of galaxy cluster formation. Finally, we summarize
and discuss our results in Section~\ref{Discussion}.

\section{Theoretical considerations} \label{Theoretical}

In the following discussion, we concentrate on viscous gases in the
astrophysical context of galaxy and galaxy cluster formation.  We briefly
review the basic physical equations governing the hydrodynamics of the
relevant class of `real' (as opposed to ideal) fluids, and the constraints
that exist for some of the free parameters that describe their properties.
This includes a discussion of internal friction processes in the collisional
regime and their relevance for astrophysical plasmas. In particular, we examine
viscous effects in intracluster gas, assuming that it is fully ionized and
that it consists of a primordial mixture of hydrogen and helium.  We also
describe the differences and difficulties that arise when magnetic fields are
present in clusters.
 
\subsection{Navier-Stokes equation}
 
To describe real fluids, two of the fundamental equations of hydrodynamics
that hold for ideal gases, namely Euler's equation and the energy conservation
law, need to be revised.  The continuity equation remains in its
familiar form, i.e.
\be
\label{continuityeq}
\frac{\partial \rho}{\partial t} + \frac{\partial(\rho v_k)}{\partial x_k} \,
= \, 0   \ee expresses mass conservation as usual, where $\rho$ is the gas
density, $v_k$ denotes the local velocity vector of the fluid, and the
summation convention has been employed.

When there is relative motion between different parts of a real fluid,
internal friction forces lead to an additional transfer of momentum that is
absent in an ideal gas, and which in general will act to reduce velocity
differences.  The friction forces modify the momentum flux density tensor,
which becomes \be
\label{momentumfluxeq}
\Pi_{ik} \,=\, p\delta_{ik} + \rho v_iv_k - \sigma_{ik}\, .
\ee
In this equation, $p$ is the gas pressure and $\sigma_{ik}$ represents the
viscous stress tensor, which to first approximation can be assumed to be a
linear function of the first spatial derivatives of the velocity field. It can be shown
\citep{LandauFM} that the most general tensor of rank two satisfying the
requested criterion is given by
\be
\label{stresstensoreq}
\sigma_{ik} \,=\, \eta \bigg( \frac{\partial v_i}{\partial x_k} +
\frac{\partial v_k}{\partial x_i} - \frac{2}{3}
\delta_{ik}\frac{\partial v_l}{\partial x_l} \bigg) + \zeta
\delta_{ik}\frac{\partial v_l}{\partial x_l} \, ,
\label{eqtensor}
\ee  
where $\eta$ is called the coefficient of shear viscosity, and
$\zeta$ represents the bulk viscosity coefficient. Bulk viscosity
becomes important when the fluid is rapidly compressed or expanded on
a timescale shorter than the relaxation time of the fluid, in which
case considerable energy can be dissipated. The coefficients of
viscosity can be functions both of gas pressure and temperature, but
not of gas velocity, because of the criterion imposed above on the
viscous stress tensor.

The generalized form of Euler's equation describing the motion of viscous
fluids can be written as
\begin{eqnarray}
\label{NavierStokeseq}
\rho \bigg( \frac{\partial v_i}{\partial t} + v_k \frac{\partial
  v_i}{\partial x_k} \bigg)  = - \frac{\partial p}{\partial x_i}
  -\rho \frac{\partial \Phi}{\partial x_i} +{}
\nonumber \\
 {}+ \frac{\partial}{\partial x_k} \bigg[ \eta \bigg( \frac{\partial
    v_i}{\partial x_k} + \frac{\partial v_k}{\partial x_i} - \frac{2}{3}
\delta_{ik}\frac{\partial v_l}{\partial x_l} \bigg) \bigg] +{}
  \nonumber \\
 {}+ \frac{\partial}{\partial x_i} \bigg( \zeta \frac{\partial
  v_l}{\partial x_l} \bigg) \, , 
\end{eqnarray}
where $\Phi$ is the  gravitational potential. When the
coefficients of shear and bulk viscosity are assumed to be constant,
equation (\ref{NavierStokeseq}) is called Navier-Stokes equation.

\subsection{General heat transfer equation}

Unlike ideal gases which are isentropic outside of shock waves, entropy
conservation does not hold for viscous fluids. In the latter case, the energy
conservation law needs to be augmented with additional terms which depend on
the viscous stress tensor and on the temperature gradient
\citep{LandauFM}. This results in 
\be
\label{Energyconservationeq}
 \frac{\partial}{\partial t} \left(\frac{1}{2} \rho v^2 + \rho \epsilon
\right)\,  = 
\, -  {\vec \nabla} \left[ \rho \vec v 
\left(\frac{1}{2} v^2 +  w \right) - \vec v \vec \sigma - \kappa \, {\vec \nabla} T \right],
\ee
where $w$ is the heat function given by $w=\epsilon + p/ \rho$, and $\kappa$ is the
coefficient of thermal conduction. Using the continuity equation and the
Navier-Stokes equation, the energy conservation law can be rewritten as
\be
\label{Generalheattransfeq}
\rho T \frac{{\rm d}S}{{\rm d}t} \, = \, {\vec \nabla} (\kappa \,{\vec \nabla}
T) + \frac{1}{2} \eta \,\sigma_{\alpha \beta} \sigma_{\alpha \beta} + \zeta
({\vec \nabla}\, v)^2 \, , 
\ee 
which is called the general heat
transfer equation. Here $\sigma_{\alpha \beta}$ denotes the shear part of the
viscous stress tensor, or `rate-of-strain tensor'. This equation expresses how
much entropy is generated by the internal friction of the gas and by the heat
conducted into the considered volume element. From the general heat transfer
equation it is evident that the coefficients of viscosity and thermal
conduction need to be positive, given that the entropy of the gas can only
increase, as imposed by the second law of thermodynamics. In the following, we
will not consider thermal conduction any further, which has recently been
discussed in independent studies that analyzed its impact on cluster
cooling flows \citep[e.g.][]{Narayan2001,Jubelgas2004,DolagJubelgas2004}.

\subsection{The viscous transport coefficients in astrophysical plasmas}

\subsubsection{Kinetic theory approach}

In the kinetic theory of neutral gases, the viscosity coefficients are
kinetic coefficients of the Boltzmann transport equation, and can be
estimated by solving this equation under the assumption that the
characteristic length-scale of the problem under consideration is much
larger than the mean free path $l$ of particles, which is the
so-called collisional regime\footnote{In this study, we will not
discuss internal friction processes in the collisionless regime,
because the relevant scales for this regime are at best partially
resolved (and often completely below the spatial resolution) in
current state-of-the-art numerical simulations of galaxy
clusters.}. From this approach it follows \citep{LandauPK} that the
shear viscosity coefficient can be expressed as \be
\label{Shearviscosityeq}
\eta \sim m\, \bar{v}\, n\, l \sim \frac{\sqrt {m \, T}} {\sigma} \, , \ee
where $ \bar{v} \sim \sqrt{T/m}$ is the mean gas velocity, $n$ is the gas
number density, and $\sigma \sim 1/(n\,l)$ is the collisional cross-section.
Equation (\ref{Shearviscosityeq}) implies that at a given gas temperature the
shear viscosity coefficient does not depend on gas pressure.  When the
Boltzmann transport equation is solved for the bulk viscosity coefficient, one
then obtains that $\zeta$ vanishes for the case of a monoatomic
non-relativistic gas.

If magnetic fields are absent in a collisional plasma, the main transfer of
momentum due to internal friction comes from the motion of ions. Hence, it is
sufficient to consider only collisions between ions, neglecting the ones
occurring with electrons, in order to estimate the amount of shear viscosity.
The cross-section in the limit of small angle scattering in the unmagnetized
Coulomb field is given by, \be \sigma_{c} \, = \, \frac{4 \pi(Ze^2)^2}{\mu^2
  |\vec{v_e}-\vec{v_i}|^4}\ln \Lambda \,, \ee where $\mu$ is the reduced mass
of electrons and ions, and $\ln  \Lambda$ is the Coulomb logarithm, which can
be approximatively taken to be 37.8 for intracluster gas \citep{Sarazin}. Under
the assumption that electrons have much higher velocity than the ions, it
follows that $\mu (\vec{v_e}-\vec{v_i})^2 \sim T_e$, for the ``e--e'' and
``e--i'' collisions, giving an expression for the mean free path of electrons
in the form 
\be 
\lambda_e \sim \frac{T_e^2}{4 \pi e^4\, n_e\, \ln \Lambda} \,.
\ee 
Similarly, the ion mean free path reads 
\be
\label{meanionpatheq}
\lambda_i \sim \frac{T_i^2}{4 \pi (Ze)^4 \,n_i\, \ln \Lambda} \,.
\ee
Thus, based on the simple derivation of the shear viscosity in the
framework of the kinetic theory (eqn.~\ref{Shearviscosityeq}) and using
the expression for  the mean free path of ions, one obtains an
estimate of the shear viscosity in the case of a fully ionized,
unmagnetized plasma \citep{LandauPK}, viz.  
\be 
\eta \sim
\frac{m_i^{1/2}\, T_i^{5/2}}{(Ze)^4\, \ln \Lambda} \,.  
\ee 
The exact magnitude of the shear viscosity coefficient is given by
\citep[e.g.][]{Braginskii1958, Braginskii1965}, 
\be
\label{Braginskiiviscosity}
\eta \, = \, 0.406 \frac{m_i^{1/2}\, (k_B \,T_i)^{5/2}  }{(Ze)^4
 \, \ln \Lambda} \,,
\ee
while the bulk viscosity coefficient remains zero. 

\subsubsection{Saturation of the viscous stress tensor}

When the length scale on which the velocity is changing becomes similar or
smaller than the mean free path of ions, an unphysical situation would occur
if the momentum transfer due to viscous forces propagates faster than
the information on changes of the pressure forces, i.e. faster then the mean sound
speed of ions \citep{Accretionpower,Sarazin}. Thus, 
internal friction forces need to saturate
at the relevant length scales
 to a strength of order of the
pressure forces. More specifically, let us define a characteristic
length-scale $l_v$ such that the shear viscous force\footnote{An analogous
  argument holds for the bulk part of the viscous force as well.} obeys \be
\label{saturation1}
F_{\rm visc} \sim \eta \frac{\sigma}{l_v} \sim \eta \frac{v_i}{l_v^2} \sim
\eta \frac{c_s}{l_v^2} \,,
\ee    
where $v_i$ is the mean velocity of ions, and $c_s$ is the sound
speed of the ions. The  criterion for viscosity saturation can be expressed as
\be
\label{saturation2}
\frac{c_s}{l_v^2} < \frac {c_s}{\lambda_i^2}\,, 
\ee
implying that if $l_v < \lambda_i$, the viscous stress tensor has to
saturate to a  value of the order of $c_s/\lambda_i$.

\subsubsection{Magnetized plasmas} \label{Magnetized_plasmas}

In the presence of magnetic fields, the viscous transport coefficients will
depend on the quantity $\omega\, \tau$, where $\omega$ is the cyclotron
frequency of the considered species (electrons or ions) and $\tau$ is the
collisional time. The transport coefficients along the field lines will have
the same form as in the case of an unmagnetized plasma, because the
charged particles can move freely along the magnetic field lines and can cover
distances of order of their mean free path. On the other hand, in the case of
strong magnetic fields, with $\omega \tau \gg 1$, the transport
coefficients will be suppressed in the perpendicular direction to the field
lines, and this suppression will be of order of $\omega  \tau$ or $(\omega
\tau)^2$ for different viscosity terms. If the temperatures of electrons and
ions are similar, as expected for intracluster plasmas, the viscosity will be
dominated by ions, even in the presence of strong magnetic fields.

It is worth pointing out that unlike to the case of an unmagnetized
plasma, where internal friction of a compressible fluid is determined
by two scalar transport coefficients, the viscous transport
coefficient becomes a tensor of rank four when there is a
non-vanishing magnetic field. Thus, given the symmetry of the viscous
transport tensor, there will in general be seven independent viscosity
coefficients, five related to the shear, and two to the bulk
flows. Therefore, the treatment of viscous flows in magnetized plasmas
becomes extremely difficult, and includes the possibility that
compressional motions provide additional viscous heating. In fact, a
plasma compression in the direction perpendicular to the magnetic
field (assuming that the field lines are ordered locally) will produce
an excess of transverse pressure, and thus will give rise to a viscous
stress with unsuppressed transport coefficient, which is known as
gyrorelaxational heating. Potentially, this heating mechanism could
operate in the case of AGN--driven bubbles that buoyantly rise in the
ICM, pushing the intracluster gas in front of them, as is seen in a
number of numerical simulations \citep[e.g.][]{Churazov2001,
Quilis2001, Hoeft2004, DallaVecchia2004, Reynolds2005, Sijacki2006}
and also   indicated by recent X--ray observations
\citep[e.g.][]{Birzan2004,McNamara2005,Nulsen2005,Fabian2006}. However, so far it
has only been possible to poorly constrain the topology of magnetic
field lines in clusters with observations, while theoretical models
offer a broad range of possible magnetic field configurations.  It
will still take some time before fully radiative MHD simulations of
galaxy clusters can overcome their present limitations and provide
clearer theoretical predictions.  Hence, it is presently difficult to
put robust constraints on the magnetic field topology in clusters, its
evolution over cosmic time, and its dependence on the dynamical state
of a cluster, even though some interesting predictions can be made
based on non-radiative MHD simulations \citep[e.g.][]{Dolag2002}. In our
numerical modelling of gas viscosity we therefore parameterize the role
of magnetic fields by introducing a suppression parameter $f$ in front
of the Braginskii viscosity. We will assume $f$ to be constant in time
and to be independent of cluster mass.
 
The discussion above is valid under the assumption that the plasma is in a
quasi steady state, where the mean values of relevant quantities change
sufficiently slowly in time and space, thus that collisions can establish a
Maxwellian distribution on the time scale $\tau$. Otherwise, field
fluctuations can significantly change the magnitude of the suppression of the
transport coefficients in the direction perpendicular to the field lines.

\section{Numerical implementation} \label{Numerical}

We use the parallel TreeSPH-code {\small GADGET-2} \citep{Gadget2,
  Springel2001} in this study, in its entropy conserving formulation \citep{SH2002}. In addition to gravitational and non-radiative
hydrodynamical processes, the code includes a treatment of radiative cooling
for a primordial mixture of hydrogen and helium, and heating by a spatially
uniform, time-dependent UV background \citep{Katz1996}.  Star
formation and associated supernovae feedback processes can also be tracked by
the code, using a simple subresolution multiphase model for the ISM
\citep{S&H2003}.

Even though the bulk viscosity is identical to zero for unmagnetized
fully-ionized plasmas, there are a number of cases where bulk
viscosity may still be important, for example in the presence of
magnetic fields where the viscosity tensor contains terms that
explicitly depend on the velocity divergence. Also for the sake of
completeness, we have therefore implemented a treatment of viscosity
in {\small GADGET-2} that accounts both for shear and bulk
viscosity. There is a small number of previous studies in the
literature that discuss SPH formalisms for internal friction processes
\citep{Flebbe1994,Schaefer2004}. Our new implementation follows a
somewhat different and more complete approach, however, and it is
consistent with the entropy-conserving formulation of SPH introduced
by \cite{SH2002}. In the following, we give a brief summary of the SPH
method, and then derive the particular formulation of the discretized
Navier-Stokes and general heat transfer equations that we adopted.

One of the central aspects of the SPH method is the idea to represent a given
thermodynamic function with an interpolant constructed from the values at a
set of disordered points. These fluid particles
are usually characterized by their position $r$, mass $m$, and
velocity $v$ \citep{Lucy1977,Gingold1977,Monaghan1985}. The computation of an
interpolant is based on a kernel function, which is often adopted as a simple
spline kernel \citep{Monaghan1985}, \be W(r,h)=\frac{8}{\pi h^3} \left\{
\begin{array}{ll}
1-6\left(\frac{r}{h}\right)^2 + 6\left(\frac{r}{h}\right)^3, &
0\le\frac{r}{h}\le\frac{1}{2} ,\\
2\left(1-\frac{r}{h}\right)^3, & \frac{1}{2}<\frac{r}{h}\le 1 ,\\
0 , & \frac{r}{h}>1 .
\end{array}
\right. \label{eqkernel} \ee where $h$ is the smoothing length. The
interpolant $\tilde Q(\vec{r})$ of a thermodynamic quantity can then
be constructed from the values $Q_i$ of the particle set as \be \tilde
Q_i = \sum_{j=1}^{N} Q_j \frac{m_j}{\rho_j} W_{ij}(h_i)\,, \ee where
the sum is evaluated over all particles, $W_{ij}(h_i)$ is an
abbreviation for $W(|\vec r_i - \vec r_j|,h_i)$, and $h_i$ is the
adaptive smoothing length of particle $i$. A derivative of the
interpolant can now be obtained straightforwardly by applying the
$\vec\nabla$-operator to the kernel function itself, viz.  
\be
\label{nablaQ} \vec\nabla_i \tilde Q_i = \sum_{j=1}^{N} Q_j
\frac{m_j}{\rho_j} \vec\nabla_i W_{ij}(h_i).  
\ee  
We use this
property of the SPH formalism to derive the viscous accelerations
exerted on gas particles. The SPH discretization of the viscous stress
tensor can be readily constructed based on standard  expressions for
velocity gradients and velocity divergence
\citep{Monaghan1992}. Specifically,  the  derivative of the
$\alpha$-component of particle $i$'s velocity with respect to
$x_{\beta}$ (where $\alpha$ and $\beta$ range from 0 to 2) can be
written as 
\be \label{SPH_velocitygradient} \frac{\partial
v_{\alpha}}{\partial x_{\beta}} \bigg|_i \,=\, \frac{1}{\rho_i}
\sum_{j=1}^{N} m_j (\vec{v_j}-\vec{v_i})\big|_{\alpha}\,
\big(\vec{\nabla}_i W_{ij}(h_i)\big)\big|_{\beta}\,.  
\ee 
Hence, the velocity divergence can be simply constructed as 
\be
\label{SPH_velocitydivergence} {\vec \nabla \cdot}\vec v_i \equiv
\frac{\partial v_{\alpha}}{\partial x_{\alpha}} \bigg|_i =
\frac{1}{\rho_i} \sum_{j=1}^{N} m_j
(\vec{v_j}-\vec{v_i})\big|_{\alpha}\, \big(\vec{\nabla}_i
W_{ij}(h_i)\big)\big|_{\alpha}\,, 
\ee 
where the summation notation for
repeated Greek indexes was adopted. Therefore, based on equations
(\ref{stresstensoreq}), (\ref{SPH_velocitygradient}) and
(\ref{SPH_velocitydivergence}), the SPH formulation of the viscous
stress tensor reads
\begin{eqnarray}
\sigma_{\alpha \beta}\bigg|_{i} = \eta \bigg( \frac{\partial
    v_{\alpha}}{\partial x_{\beta}} \bigg|_i + \frac{\partial
    v_{\beta}}{\partial x_{\alpha}} \bigg|_i -
\frac{2}{3}\delta_{\alpha \beta} \frac{\partial v_{\gamma}}{\partial x_{\gamma}} \bigg|_i \, \bigg) \, +{}
\nonumber \\
{}+ \, \zeta \, \delta_{\alpha \beta} \frac{\partial
    v_{\gamma}}{\partial x_{\gamma}} \bigg|_i\,.
\end{eqnarray} 
Considering equation (\ref{NavierStokeseq}) and using the notation
introduced in equation (\ref{nablaQ}), the following expression for
the acceleration of gas particles due to the shear forces can be
readily derived
\begin{eqnarray}
\label{shearacc}
\frac{{\rm d}\vec v}{{\rm d}t}\bigg|_{i,\rm shear}\! = \, \sum_{j=1}^{N} m_j \, \bigg[ \frac{\eta_i
    \vec{\sigma}_i}{\rho_i^2}\, \vec{\nabla}_i W_{ij}(h_i)\, +{}
\nonumber \\
{}+ \,\frac{\eta_j
    \vec{\sigma}_j}{\rho_j^2} \, \vec{\nabla}_i W_{ij}(h_j) \bigg] \, , 
\end{eqnarray}
where the product of $\eta$ and $\vec\sigma$ gives the shear
part of the viscous stress tensor, or in the other words,
$\vec{\sigma}_i$ is now the rate-of-strain tensor of particle
$i$. The previous equation can be
  written in an explicit component form as follows
\begin{eqnarray}
\label{shearacc_comp}
\frac{{\rm d} v_\alpha}{{\rm d}t}\bigg|_{i,\rm shear}\! = \, \sum_{j=1}^{N} m_j \, \bigg[ \frac{\eta_i
    {\sigma}_{\alpha \beta}|_i}{\rho_i^2}\, \big(\vec{\nabla}_i W_{ij}(h_i)\big)\big|_\beta\, +{}
\nonumber \\
{}+ \,\frac{\eta_j
    {\sigma}_{\alpha\beta}|_j}{\rho_j^2} \, \big(\vec{\nabla}_i W_{ij}(h_j)\big)\big|_\beta \bigg] \,. 
\end{eqnarray}
We note that this equation conserves linear momentum, but does
not manifestly conserve the total angular momentum. A non-conservation
of angular momentum can arise due to the fact that even though the
force is clearly antisymmetric, the viscous stress tensor can induce
torques, and thus the force between two particles is not necessarily
central any more. Note that this is a consequence of the tensor nature
of the viscous stresses, and not an artificial feature of our
numerical scheme. In order to circumvent this apparent inconsistency,
one could introduce an additional intrinsic property for every gas
particle, namely a spin variable, that would store how much torque has
been exerted on it and would itself be a source of shear between two
particles which would try to keep this spin close to zero. However, the
non-conservation of angular momentum due to viscous forces in our
formulation is basically negligible, as has already been discussed in
detail in \cite{Riffert1995}. Comparing the discretized form of the
SPH equations to the continuum limit shows that the angular momentum
is conserved to an accuracy of order ${\cal O}(h^2)$, which is
comparable to the
error made in the usual SPH kernel estimates of other fluid
quantities, like the density. In an analogous manner to equation (\ref{shearacc}), the
acceleration caused by forces due to bulk viscosity can be estimated
as
\begin{eqnarray}
\label{bulkacc}
\frac{{\rm d}\vec v}{{\rm d}t}\bigg|_{i,\rm bulk}\!  = \, \sum_{j=1}^{N} m_j \, \bigg[
  \frac{\zeta_i \, \vec{\nabla} \cdot \vec v_i}{\rho_i^2} \, \vec{\nabla}_i W_{ij}(h_i) \,
  +{}
\nonumber \\
{} + \, \frac{\zeta_j \, \vec{\nabla} \cdot \vec v_j}{\rho_j^2} \, \vec{\nabla}_i
  W_{ij}(h_j) \bigg] \, .
\end{eqnarray} 

We employ the specific entropy of an SPH particle as independent thermodynamic
variable. Instead of using the conventional
thermodynamic entropy directly, it is however more convenient 
to replace the entropy $S$  with an
entropic function $A$, related to the entropy by
\be
\frac{{\rm d}A(S)}{{\rm d}t} \,=\, \frac{\gamma -1}{\rho^{\gamma -1}}\, T
\,\frac{{\rm d}S}{{\rm d}t} \,,
\ee
where $\gamma$ is the adiabatic gas index. Therefore, from the
general heat transfer equation it follows that the increase of the
entropic function due to  internal friction forces
is given by
\bea
\label{shearvisc}
\frac{{\rm d}A_i}{{\rm d}t} \bigg|_{\rm shear} &=& \, \frac{1}{2}
\,\frac{\gamma
  -1}{\rho_i^{\gamma -1}} \,\frac{\eta_i}{\rho_i} \,\sigma_i^2 \\
\frac{{\rm d}A_i}{{\rm d}t} \bigg|_{{\rm bulk} \enskip} &=& \, \frac{\gamma
  -1}{\rho_i^{\gamma -1}}\, \frac{\zeta_i}{\rho_i}\, ({\vec \nabla \cdot}\vec
v_i)^2. 
\eea 
Clearly, this formulation shows that the entropic function can
only increase due the action of internal friction forces if the shear and bulk
viscosity coefficients are positive, as desired.

We have implemented different parameterizations of the viscosity coefficients
in the simulation code. Besides a model with constant viscosity, we realized a
model for cosmological applications where the shear viscosity is parameterized
with equation (\ref{Braginskiiviscosity}), modified with an additional
prefactor that controls in a simple way the possible shear viscosity
suppression due to the presence of magnetic fields, as discussed in
Section~\ref{Magnetized_plasmas}. We follow the literature and vary this
prefactor in the range of $0.1$ to $1.0$.  We have also introduced an
additional time-step criterion in the code in order to protect against
situations where the Courant timestep may not be small enough to guarantee
accurate integration of large viscous stresses. We have adopted the maximum
allowed time-step as \be {\rm d}t_{\rm max} \le {\rm d}t_{\rm visc} \quad {\rm
  with} \quad {\rm d}t_{\rm visc} = \alpha \frac{A}{|\dot A_{\rm visc}|}\,,
\ee where $\dot A_{\rm visc}$ represents the rate of increase of the entropic
function due to shear and bulk viscous forces, and $\alpha$ is a dimensionless
time-step parameter that controls the integration accuracy.

Finally, we estimate the characteristic length-scale $l_v$ in the code.
Following the arguments expressed in equations (\ref{saturation1}) and
(\ref{saturation2}), we adopt a saturation of the relevant viscous stress
tensor components to a value given by $\sim c_s/\lambda_i$, provided $l_v$ is
found to be smaller than the ion mean free path.

  At the end of this Section, we briefly discuss the differences
  that exist between the functional forms of the physical and the
  artificial viscosity, bearing in mind their conceptual
  differences. As discussed in detail in \cite{Gadget2}, the {\small
  GADGET-2} code computes the acceleration due to the artificial
  viscosity as follows \be \frac{{\rm d} \vec v_i}{{\rm d}t}
  \bigg|_{visc} = - \sum_{j=1}^{N} m_j \, \Pi_{ij} \vec{\nabla}_i
  \overline{W}_{ij} \,, \ee where $\overline{W}_{ij}$ represents the
  arithmetic mean of $W_{ij}(h_i)$ and $W_{ij}(h_j)$. The entropy
  increase due to the action of artificial viscous forces is given by
  \be \frac{{\rm d}A_i}{{\rm d}t} = \frac{1}{2}
  \frac{\gamma-1}{\rho_i^{\gamma -1}} \sum_{j=1}^N m_j \, \Pi_{ij} \,
  \vec v_{ij} \cdot \vec{\nabla}_i \overline{W}_{ij} \,, \ee with
  $\vec v_{ij} \equiv \vec v_i - \vec v_j$, and $\Pi_{ij}$ is defined
  in a slightly different form\footnote{For a more sophisticated
  form of the artificial viscous term see \cite{Cleary1998} and
  \cite{ClearyHa2002}.}  \citep{Monaghan1997} compared to the one that
  is usually adopted in many SPH codes \citep{Monaghan1983}, namely
  \be \Pi_{ij} = - \frac{\alpha}{2} \frac{(c_i + c_j -
  3w_{ij})\,w_{ij}}{\rho_{ij}}\,, \ee if the particles are approaching
  each other, otherwise $\Pi_{ij}$ is set to zero. Here $\alpha$ is a
  parameter that regulates the strength of the artificial viscosity,
  $c_i$ and $c_j$ are the sound speeds of particles $i$ and $j$,
  respectively, and $w_{ij} = \vec v_{ij} \cdot \vec r_{ij} / |\vec
  r_{ij}|$. In addition, following the arguments explained in
  \cite{Balsara1995} and \cite{Steinmetz1996}, the strength of the
  artificial viscosity is reduced in the presence of local shear in
  order to avoid spurious angular momentum transport. Comparing the
  equations for acceleration and entropy generation due to artificial
  viscous forces with the analogous expressions for the case of
  physical viscosity, it can be seen that the dependences on gas
  velocity and temperature are quite different. In particular, the
  Braginskii-parameterization of the shear viscosity coefficient has a
  much stronger dependence on gas temperature than the artificial
  viscosity, meaning that the relative importance of the physical
  viscosity is expected to be be different for objects of different
  virial temperature. It should also be stressed that the artificial
  viscosity becomes relevant only when particles are approaching each
  other, while that is not the case for the physical
  viscosity. Furthermore, given that the artificial viscosity is
  suppressed in the presence of significant local shear, the
  artificial viscosity cannot mimic the behavior of physical shear
  viscosity, as we explicitly confirmed with a number of test
  problems that will be discussed in the next Section.

\section{Illustrative test problems} \label{Illustrative}

The purpose of this section is to test the validity and applicability of our
new viscosity implementation. To this end, we consider simple hydrodynamical
problems with known analytic solutions for real (viscous) fluids. We will
first discuss the motion of an incompressible fluid under the action of shear
forces in two different situations that illustrate the typical behaviour of
viscous fluids. At the end of this section, we then compare the behaviour of
shock tube tests when physical shear and bulk viscosity are used to capture
shocks instead of the standard artificial viscosity.

We begin our investigation with the problem of a flow between two
moving planes with a finite separation $h$.  While there always exists
a flow solution for every particular initial condition, it is
interesting to note that it is not guaranteed that the solution will
be steady and stable for a different internal viscosity value. In
fact, for the case of an ideal fluid the flow is actually always
unstable, because any small perturbation in the flow will typically
not be damped in this case but rather grow in time. However, stability
of the flow can be recovered if the fluid has a sufficiently small
Reynolds number, given by \be {\cal R} \,=\, \frac {\rho u l}{\eta}
\,=\, \frac{u l}{\nu}\,, \ee where $u$ is the characteristic velocity
of the problem, $l$ is its characteristic length-scale, and $\nu \,=\,
\eta / \rho$ is the kinematic viscosity. It follows that in order to
ensure a laminar flow in the `pipe' between the two planes, the mean
velocity times the diameter of the pipe should be of the order of the
kinematic viscosity. This condition is satisfied in our numerical
tests.

\subsection{Flow between two sheets with a constant relative velocity}

As a first test we simulate the elementary hydrodynamical problem of the
motion of a viscous, incompressible fluid between two infinite parallel planes
spaced a distance $h$ apart. The space between the planes is uniformly filled
with a fluid of constant pressure, having a fixed amount of shear viscosity
and bulk viscosity equal to zero. The planes move with a constant relative
velocity with respect to each other (along the $x$-axis, for definiteness),
while the fluid is initially at rest. We expect that after a brief time
interval a stationary solution should be established, with a laminar flow
where all relevant quantities depend only on the position $y$ along the axis
orthogonal to the planes.  Solving the Navier-Stokes equation for this problem
yields that the $x$-component of the gas velocity should be a linear function
of $y$, with a  slope and zero point such that the boundary conditions at the
planes are matched, i.e.~here the fluid velocity will  be equal to the
velocity of the planes themselves. If the boundary conditions
are given by $v_x(0) = u_1$ and $v_x(h) = u_2$, then the gas velocity is simply
\be 
v_x(y) = \frac{u_2 -u_1}{h} y + u_1 \,.
\ee
The only non-zero component of the viscous shear stress tensor
is a linear function of the velocity gradient along the $y$-axis, namely
\be
\label{sigmaxy_test1}
\sigma_{xy} = \eta \frac{\partial v_x}{\partial y} = \eta \frac{u_2 -
  u_1}{h} \,.  
\ee 
Note also that it is directly proportional to the shear viscosity
coefficient, allowing us to validate whether the level of viscosity acting in
our numerical simulations actually matches the one we intended to put in.

\begin{figure*}
\centerline{
\hbox{
\psfig{file=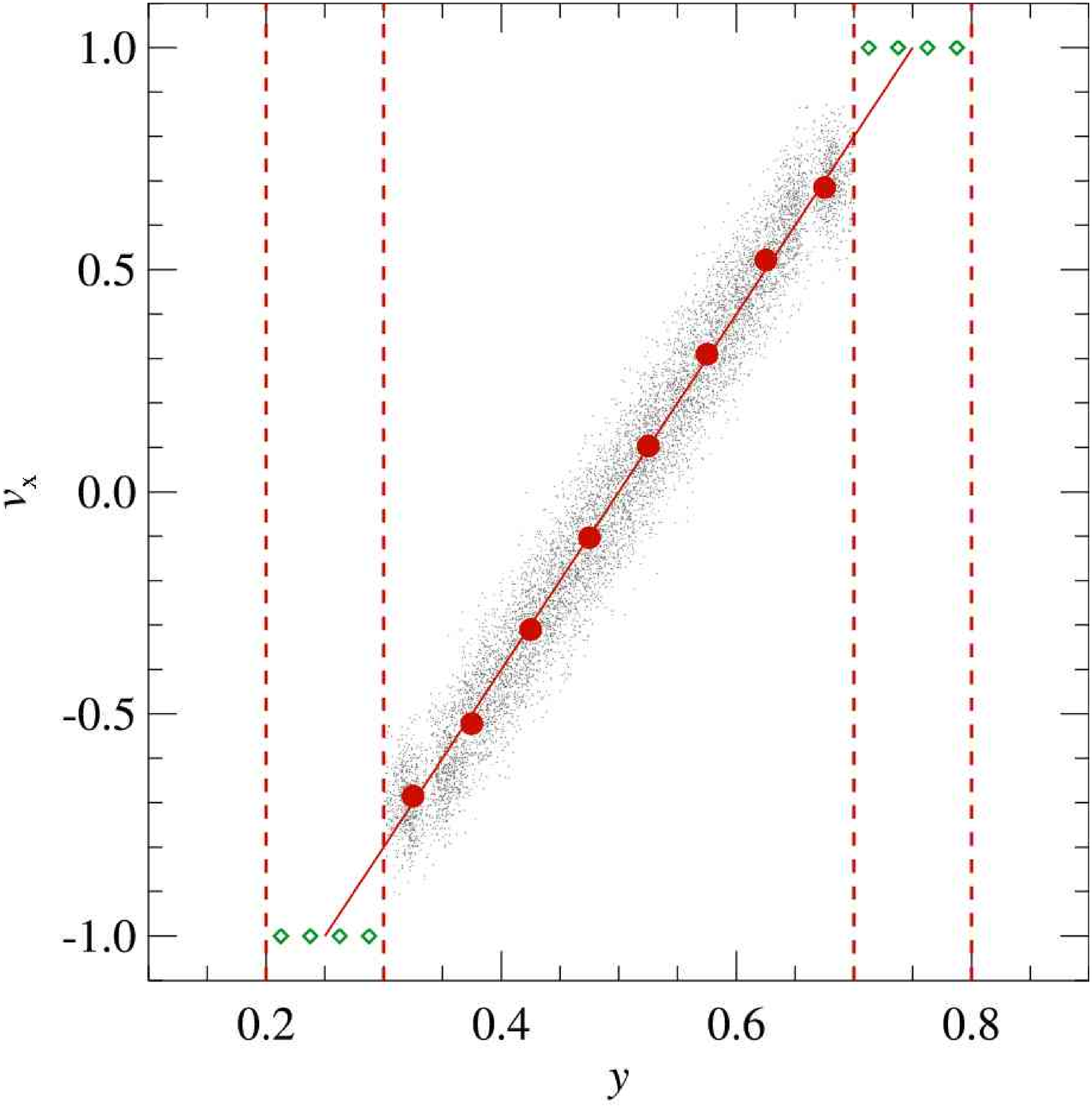,width=8truecm,height=7.5truecm}
\hspace{0.3truecm}
\psfig{file=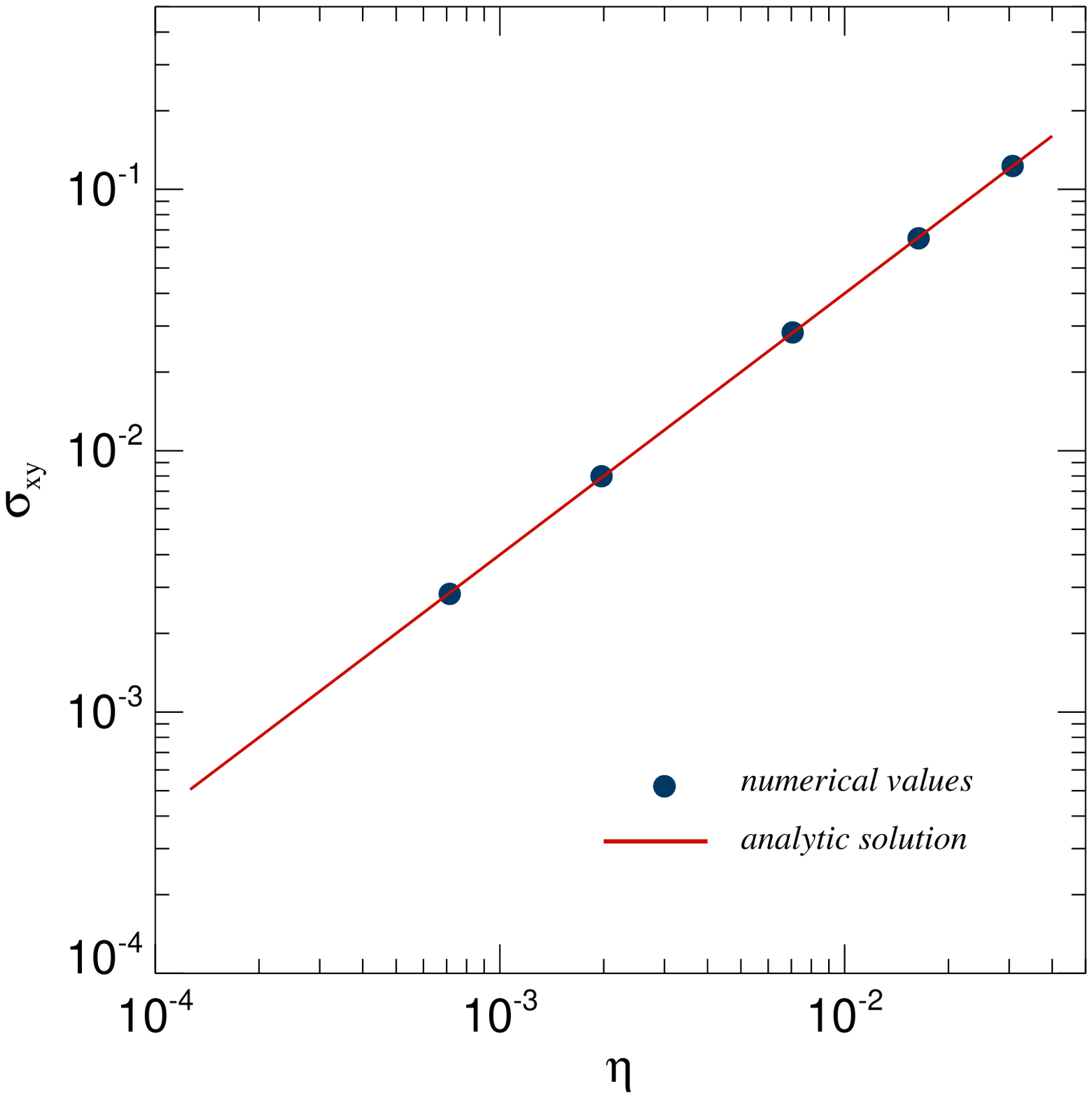,width=8truecm,height=8truecm}
}}
\caption{The panel on the left shows the stationary-state velocity profile of
  a viscous flow in between two infinite planes that move relative to each
  other.  The small grey dots represent individual gas particles in the
  simulation (only every 25th particle has been plotted for clarity). The big
  red dots are mean $v_x$-values evaluated in equally sized $y$-bins. The
  green diamond symbols show particles that belong to the two thin layers used
  to impose the boundary conditions, while the continuous solid line is the
  analytic solution. This run has been performed assuming a shear viscosity
  coefficient of $\eta \sim 0.002$ in internal code units. The mean value of
  the viscous stress tensor, $\sigma_{xy}$, as a function of the shear
  viscosity coefficient for a number of similar runs is illustrated in the
  right panel. In each case, the mean value of $\sigma_{xy}$ has been
  estimated once the flow has reached a stationary state and is plotted with a
  filled circle. The continuous line gives the analytic expectation.}
\label{Test1}
\end{figure*}

In order to simulate this hydrodynamical problem we have set up initial
conditions using a three-dimensional periodic grid with equally spaced gas
particles, all of equal mass and pressure, and being initially at rest. The
aspect ratio of the box was shorter in the $y$-direction. The motion of the
two planes was imposed by treating the particles in two thin sheets adjacent
to the planes as 'boundary particles', giving them the velocity of the
corresponding plane, and preventing them from feeling hydrodynamical forces,
i.e.~they always keep moving with their initial velocity. 

When the simulation is started, the $x$-component of the gas velocity develops
a linear dependence on $y$ under the action of the shear viscosity, and soon the
flow becomes stationary. The time needed to reach the stationarity depends on
the amount of shear viscosity. The more viscous the medium, the sooner the flow
reaches the steady state. Note that the same behaviour cannot be obtained
with the standard artificial viscosity. Also, the presence of some amount of
artificial viscosity besides the given shear viscosity perturbs the linear
dependence of the velocity on the $y$-coordinate.

In the left panel of Fig.~\ref{Test1}, we show the $x$-component of the gas
velocity as a function of $y$ when the flow has reached stationarity. The grey
little dots represent individual gas particles, while the red big dots denote
the mean $v_x$ evaluated in equally sized $y$-bins.  The solid line is the
analytic solution, while the diamonds denote particles that are part of the
boundary layers of the finite dimension of $0.1$, one moving with a velocity
$v_x = -1$, the other with $v_x = 1$. It can be seen that the numerical result is
reproducing the analytic solution with good accuracy. Note that the gas
velocity near the planes cannot reach the theoretically expected value,
because it is here fixed to the value prescribed for the two boundary layers.
A finite width of these layers is necessary to impose the boundary conditions
in a numerically robust way, but by using a larger particle number, the thickness
of this region could be made arbitrarily small, if desired. In the right
panel of Fig.~\ref{Test1}, we show the mean value of the $xy$-component of the
viscous stress tensor as a function of the shear viscosity coefficient adopted
in a specific run. The numerical values for the stress tensor have been
evaluated once the flow has reached a stationary state. The filled circles give
the mean value of $\sigma_{xy}$ for the different runs, while the solid line
is the analytic fit. It can be seen that the analytic solution is recovered
with high accuracy for a significant range of shear viscosities.

\subsection{Flow between two planes with a constant gravitational acceleration}

Another elementary hydrodynamic problem involves the viscous flow of a fluid
between two planes under the action of a constant gravitational acceleration.
This problem is equivalent to the classic example of a flow with a constant
pressure gradient \citep{LandauFM}. The initial situation is quite similar to
the previous problem, but this time the planes do not move with respect to
each other. However, there is a constant gravitational acceleration acting
along the $x$-axis. Again, we consider an incompressible fluid, so that for
symmetry reasons all quantities depend only on $y$ if a stationary laminar
flow develops. The velocity is expected to exhibit a characteristic quadratic
dependence on $y$, of the form 
\be 
v_x(y) = - \frac{\rho}{2 \eta} \frac{{\rm d}
  \Phi}{{\rm d}x} y^2 + c_1 y + c_2 \,, 
\ee 
where $\rho$ is the gas density,
$\Phi$ is the gravitational potential, and $c_1$ and $c_2$ are two constants
defined by the boundary conditions. Again, the only non-trivial component of
the viscous shear tensor is $\sigma_{xy}$, and it is related to the velocity
field in the same way as in equation (\ref{sigmaxy_test1}).

\begin{figure}
\bc
\centerline{\includegraphics[width=8.5truecm,height=8.5truecm]{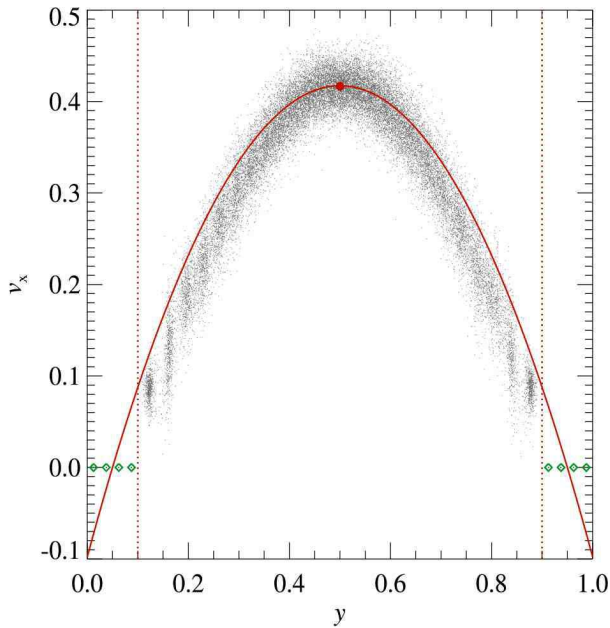}}
\caption{Velocity profile for the viscous flow between two fixed planes in a field of
  constant gravitational acceleration.  The little grey dots are the
  individual particles from the simulation output (only every 25th particle
  has been plotted for clarity).  The analytic solution with its
  characteristic quadratic velocity dependence on the $y$ coordinate is shown
  with a red solid line. The big red dot marks the mean velocity of gas
  particles for $y=0.5$, while the green diamond symbols represent the
  particles that belong to the layers of particles used to impose the boundary
  conditions.}
\label{Test2}
\ec
\end{figure}

The initial conditions for a numerical model of this problem were set
up as before, except that a constant gravitational acceleration along
the $x$-axis was imposed.  All particles were initially at rest, and
the particles of the two boundary layers were made to ignore the
gravitational field so that their positions stayed fixed.  In
Fig.~\ref{Test2}, we show a measurement of the velocity of the gas
particles once the flow reached a stationary state.  The grey little
dots are individual particles in the simulated region between the
planes, while the green diamond symbols represent particles of the
boundary layers.  The solid line gives the analytic solution, while
the big red dot shows the average velocity of gas particles for
$y=0.5$.  The central part of the flow matches the characteristic
quadratic form of the analytic solution accurately, with a maximum gas
velocity corresponding closely to the analytic solution, even though
the simulated gas velocity near the planes is a bit lower than the
analytic expectation. Again, the latter effect is to be expected due
to the finite width of the boundary layers, which also influences the
properties of the flow in their immediate vicinity.

We note that the magnitude of the velocity scatter of individual particles
around the mean profile depends on the strength of the adopted gas pressure
relative to the viscous forces. Even for parameter choices where this scatter
becomes large, the mean velocity field tracks the analytic solution well in
all cases we examined, indicating that our implemented scheme is quite robust.

\subsection{Shock tube tests}

In this section, we examine whether our new implementation of physical
viscosity can also be used to capture shocks, and how this fares with respect
to results obtained with the standard artificial viscosity. For this purpose,
we performed a number of shock tube simulations which have been setup
following the standard approach outlined in \cite{Sod}. We considered both
mild shocks with a Mach number of order 1.5, and also stronger shocks up to a
Mach number of 10. These tests allow us to constrain the amount of physical
shear and/or bulk viscosity needed to capture the shocks accurately, and hence
to assess if and how much artificial viscosity is still required in
simulations of viscous gases.

Our initial conditions consist of a three-dimensional periodic box that is
elongated in the $x$-direction, with a total length $L$.  The box is filled
with gas particles of equal mass arranged on a grid.  The left half of the box
($x < L/2$) has a higher initial pressure with respect to the right half ($x >
L/2$), such that shocks of different strength can be driven into the right
side, depending on the initial pressure ratio. The adopted adiabatic index is
$\gamma=1.4$. All gas particles are initially at rest, and we have evolved the
simulations until a final time of $t=3.5$.

Before we discuss the results it should be noted that there is an important
conceptual difference between simulations with our new implementation of
internal friction and simulations that use the artificial SPH viscosity. In
the former case we consider real gases which have different hydrodynamical
properties in the region of shocks (where the velocity field has strong
gradients) than ideal fluids for which the analytic solutions of shock tubes
refer to.  In order to properly treat real fluids in shocks, one in principle
needs to invoke the kinetic theory of gases, because the mean free path of
particles is of the order of the shock width.  This is beyond the scope of
this work. However, the analytic solutions for ideal gases outside the shock
region provide a very good approximation because the viscous forces there are
negligible, as we will explicitly confirm below.

\begin{figure*}
\centerline{
\vbox{
\psfig{file=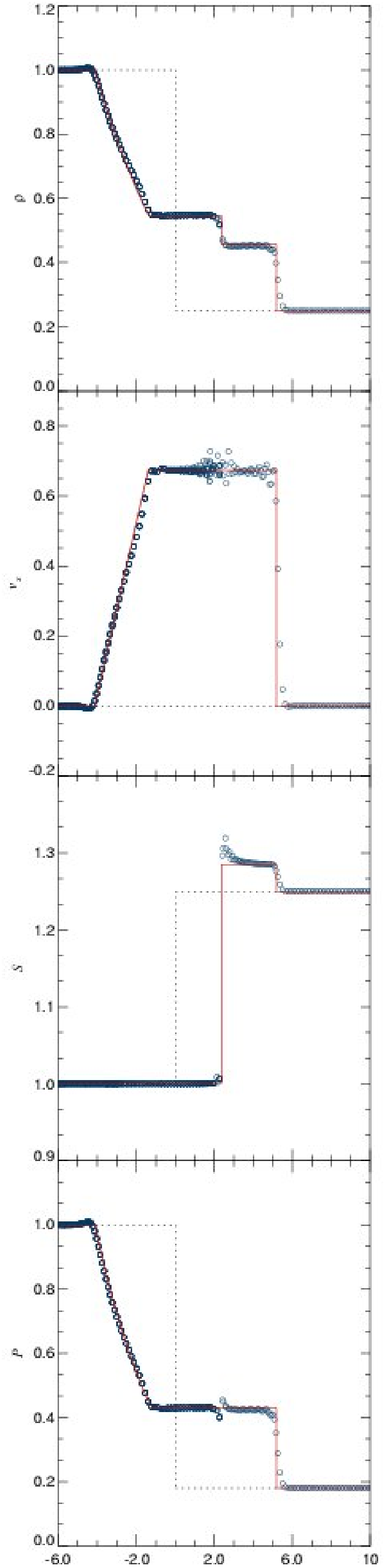,width=5.5truecm,height=20truecm}
\psfig{file=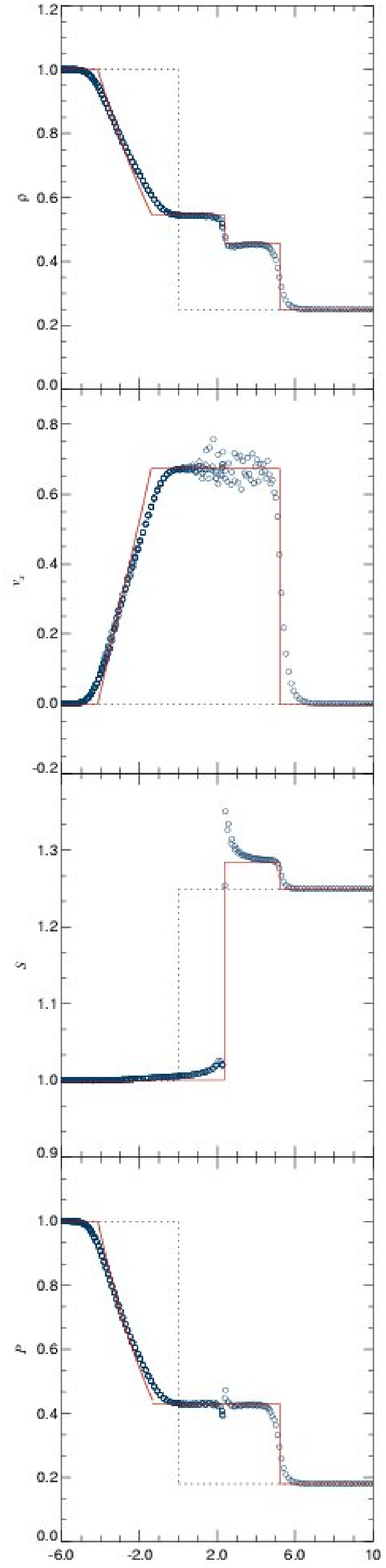,width=5.5truecm,height=20truecm}
\psfig{file=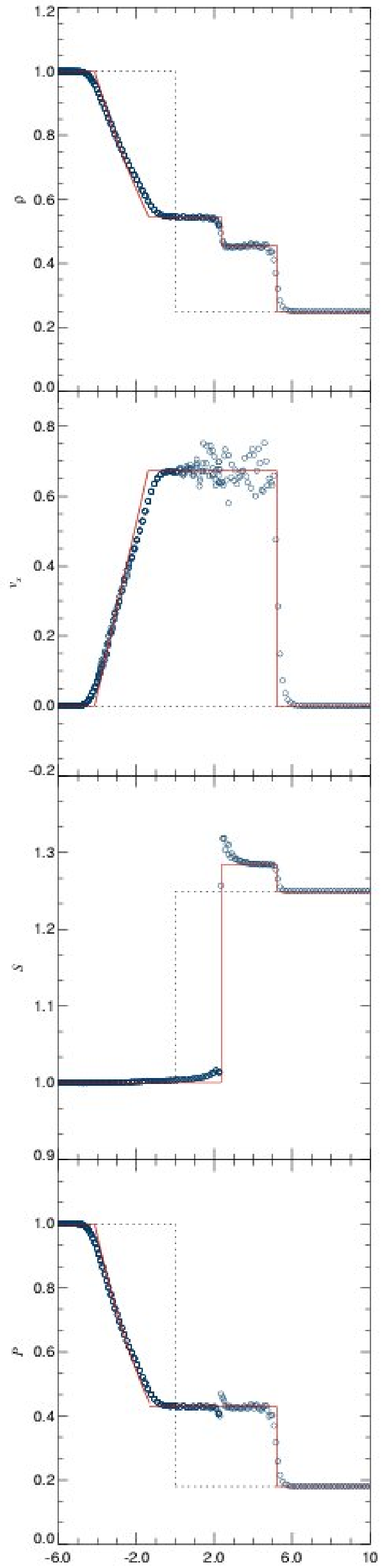,width=5.5truecm,height=20truecm}
}}
\caption{Hydrodynamical properties along a shock tube simulation
  with a shock of Mach number ${\cal M} = 1.48$, at time $t=3.5$. The initial
  conditions are drawn with doted lines, the analytic solutions are shown
  with continuous red lines, and the symbols give the SPH result. The three
  vertical columns refer to a run carried out with the standard artificial
  viscosity (left column), to one with physical shear viscosity instead
  (middle column), or to one with physical bulk viscosity (right column). From
  top to bottom, the individual rows show the profiles of density, velocity,
  entropy, and pressure, respectively.}
\label{Mach1.48}
\end{figure*}

In Fig.~\ref{Mach1.48}, we show the profile of gas density, velocity, entropy
and pressure in a shock tube calculation where the gas particles experience a
shock of strength ${\cal M}=1.48$.  The simulation results are represented by
blue circles, the dotted lines denote the initial conditions, and the
continuous red lines give the analytic solution, obtained by solving the
hydrodynamical equations of an ideal gas with imposed Rankine-Hugoniot
conditions \citep[e.g.][]{Courant,RasioShapiro91}. The three different columns
give results for the standard artificial viscosity with $\alpha= 0.7$ (left
panel), physical shear viscosity with $\eta=0.04$ (middle panel), and physical
bulk viscosity of $\zeta=0.03$ (right panel). The only difference between
these three runs lies in the gas viscosity, all the other code parameters and
the initial conditions were kept exactly identical in order to facilitate a
clear comparison. Fig.~\ref{Mach1.48} shows that the numerical model for
physical viscosity is capable of capturing the shock, and it results in quite
accurate estimates of the post-shock quantities. This holds both for shear and
bulk viscosity. Compared to the case with an artificial viscosity, there is
more velocity noise in the postshock region, however.  Also, the shock front
itself is sharper when an artificial viscosity is used, and the analytic
solution for the rarefaction wave is recovered more accurately in the
transition region to the constant density sections of the flow.  In general,
the physical viscosity solutions appear more smoothed in the transition
regions between the different parts of the flow.

\begin{figure*}
\bc
\centerline{\includegraphics[width=15truecm,height=5truecm]{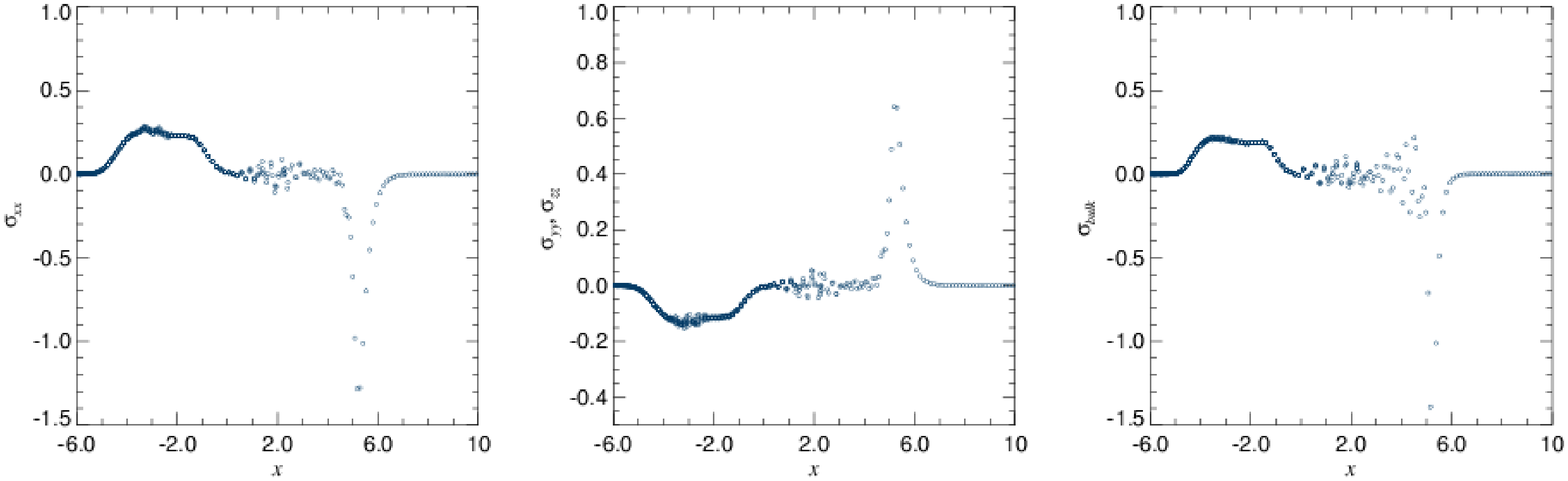}}
\caption{Viscous stress tensor profile in the case of a ${\cal M} = 1.48$ shock
  tube simulation. The first two panels refer to the physical shear viscosity,
  showing diagonal components of the stress tensor, while in the right panel
  the bulk stress tensor is plotted. In the middle panel, the dots give only
  results for $\sigma_{yy}$, because the $\sigma_{yy}$ and $\sigma_{zz}$
  components of the shear tensor show a practically identical dependence on
  the $x$-coordinate, given the symmetry of the problem.}
\label{Mach1.48_sig}
\ec
\end{figure*}

In Fig.~\ref{Mach1.48_sig}, we examine the viscous stress tensor of gas
particles in this problem.  The first two panels show the diagonal components
of the shear stress tensor, while the last panel gives the bulk viscosity
tensor. The viscous stress tensor is different from zero only in the region
between the head of the rarefaction wave and the shock wave, implying that the
viscous forces are important in that region and are negligible elsewhere. The
$\sigma_{yy}$ and $\sigma_{zz}$ components of the shear tensor behave almost
identically because of the symmetry of the problem, and thus in the middle
panel of Fig.~\ref{Mach1.48_sig} the dots are for $\sigma_{yy}$ only. It can
be noted that $\sigma_{xx}$, $\sigma_{yy}$ and $\sigma_{zz}$ have sign and
magnitude such that their sum is zero to high accuracy, as it should be given
that the shear tensor is traceless. Also, the off-diagonal components of the
shear stress tensor are negligible, as expected. The bulk stress tensor shows
very similar features as the $\sigma_{xx}$ component of the shear tensor, due
to the fact that the dominant term in both cases is $\partial v_x/\partial x$.
However, $\sigma_{\rm bulk}$ shows more scatter for $x \in [3,5]$ because the
noise in the remaining velocity derivatives in the corresponding simulation is
larger. In our simulations with with larger Mach numbers, we obtained
qualitatively very similar results as the ones presented in
Fig.~\ref{Mach1.48_sig}.

The above analysis has shown that both shear and bulk physical viscosity are
in principle capable of capturing shocks, provided the viscosity coefficients
are sufficiently large. This means that in simulations of low Reynolds number
one can probably avoid the use of any additional artificial viscosity.  In
general however, it seems clear that an artificial viscosity is still needed
even when physical viscosity is modelled. This is simply because the strength
of the physical viscosity can be quite low, and can vary locally with the flow
if a physical parameterization like that of Braginskii is used. Without
artificial viscosity, shocks would then not be captured accurately in a narrow
shock front, and particle interpenetration would not be properly avoided.
Instead, strong fluid instabilities could develop in the shock region,
growing to such large enough size that the residual physical viscosity can
damp them out. In the rest of our study, we will therefore invoke when
necessary an additional artificial viscosity in the standard way when
we model physical viscosity in astrophysically interesting
situations. This guarantees that shocks are always captured equally
well as in standard SPH.

\begin{figure*}
\centerline{
\hbox{
\psfig{file=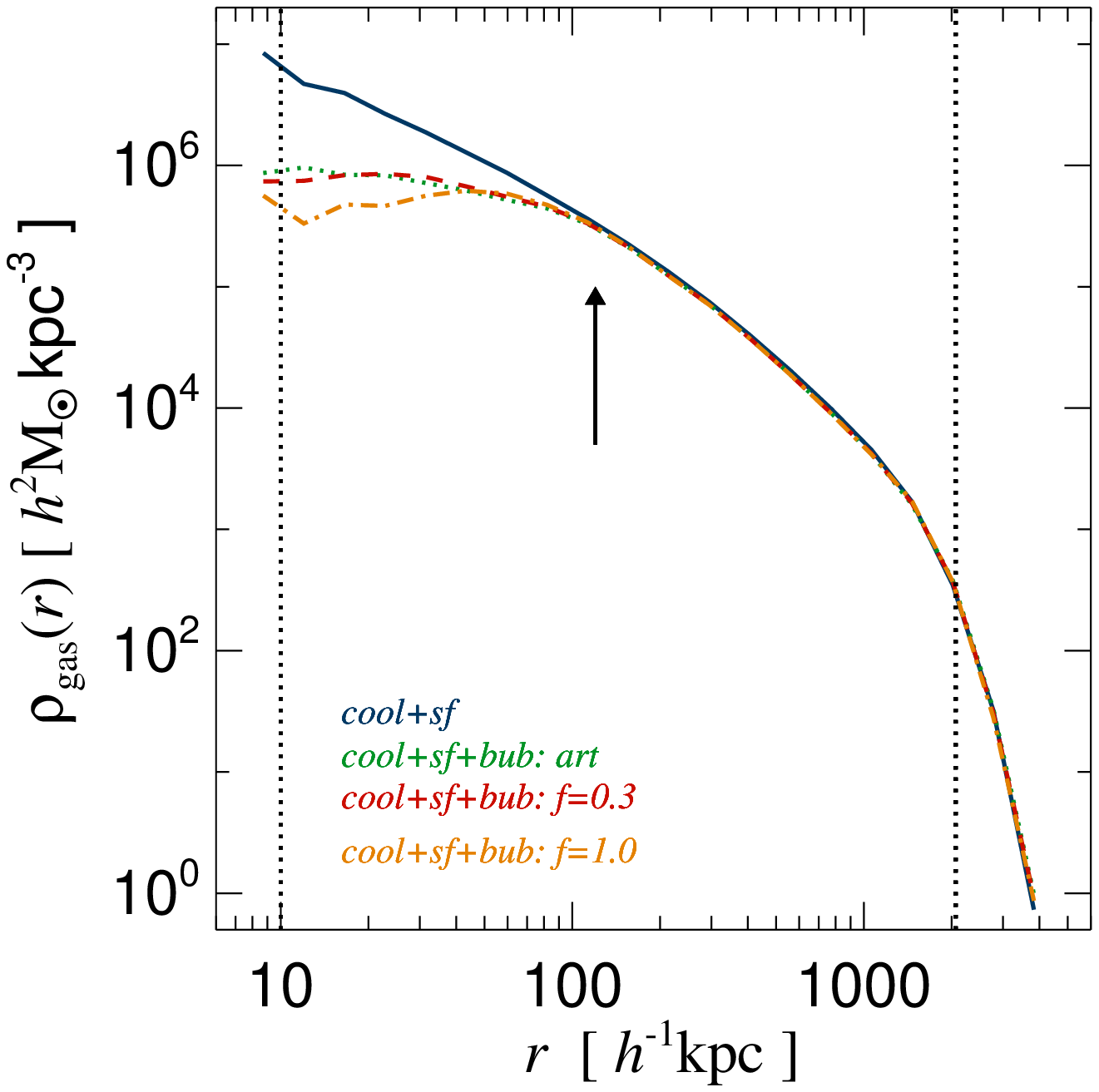,width=6truecm,height=6truecm}
\psfig{file=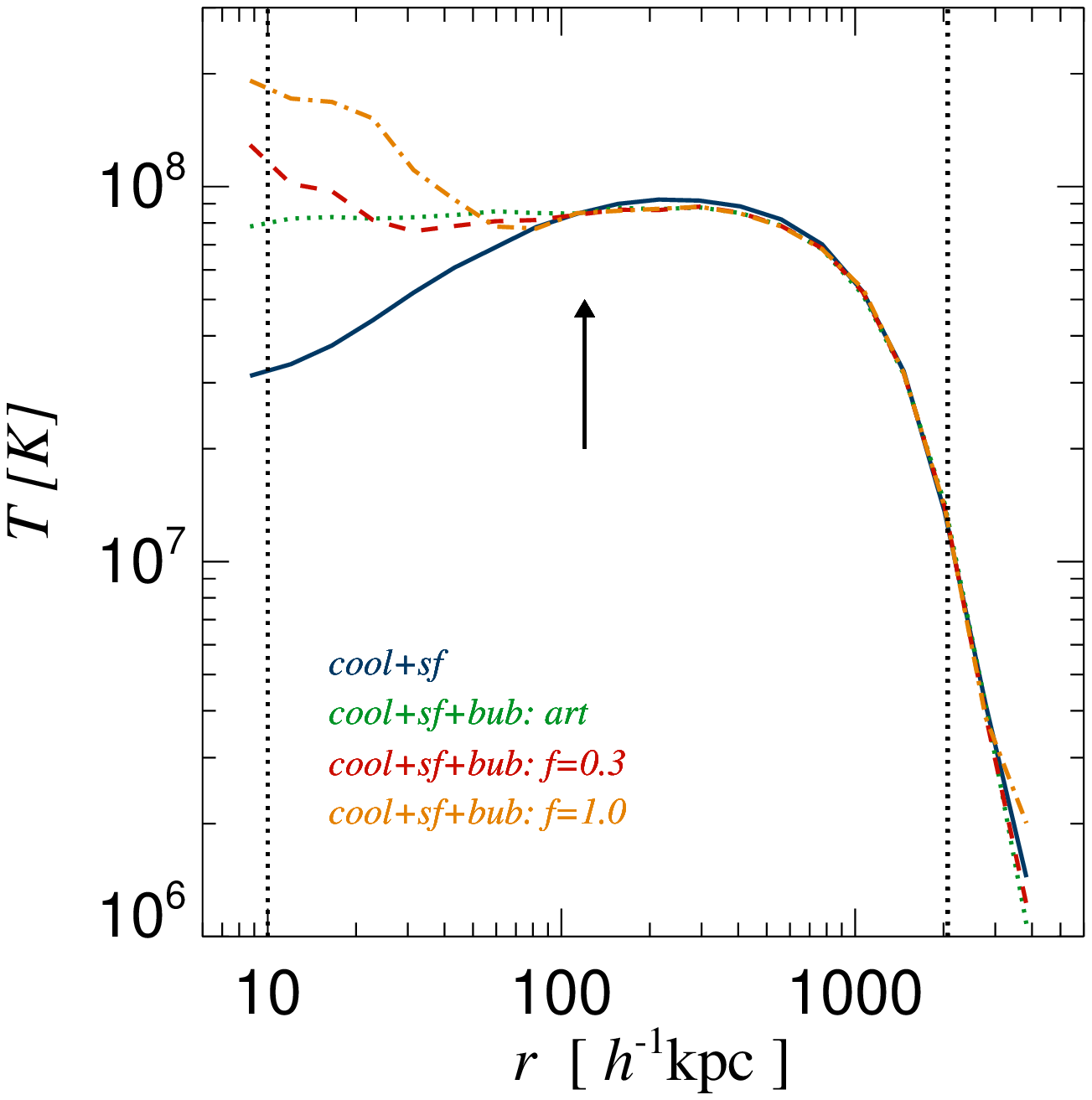,width=6truecm,height=6truecm}
\psfig{file=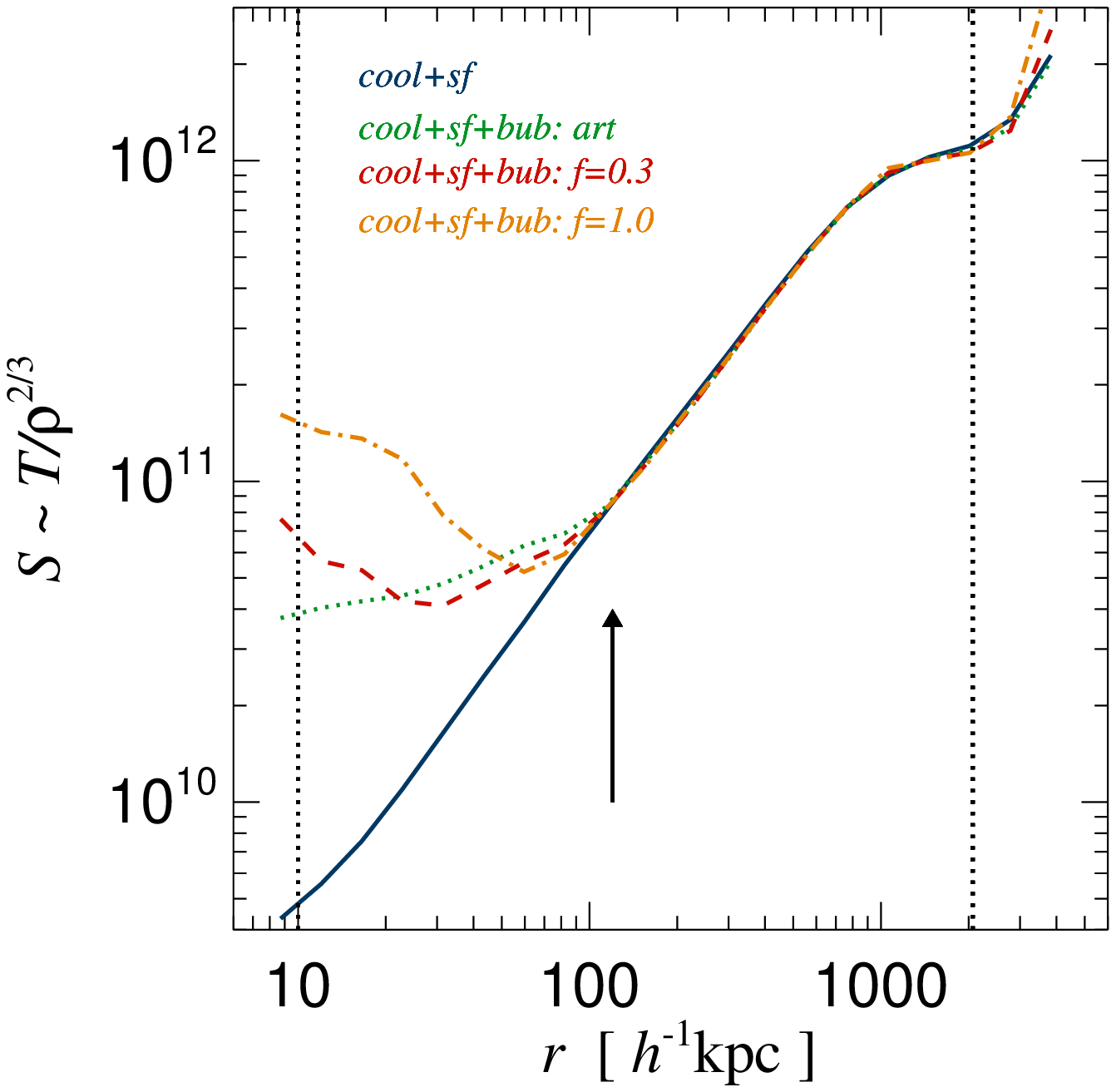,width=6truecm,height=6truecm}
}}
\caption{Radial gas profiles of a $10^{15}\,h^{-1} {\rm M_\odot}$ isolated
  halo at time $t=0.15\,t_{\rm Hubble}$. We show the gas density (left panel),
  mass-weighted temperature (middle panel), and gas entropy (right panel), and
  compare runs with cooling and star formation only (blue solid lines) with
  runs having additional AGN heating as well. The green dotted lines are for
  simulations with standard artificial viscosity.  The red dashed lines give
  results when the Braginskii parameterization of the shear viscosity is
  ``switched on'', with a suppression factor of $0.3$. For comparison, the
  orange dot-dashed lines show the results when no suppression factor is used
  for the shear viscosity.}
\label{profiles_iso}
\end{figure*}

\section{AGN--driven bubbles in a viscous intracluster medium} \label{AGN--driven}

In this section, we study the interaction of AGN-induced bubbles with a
viscous intracluster medium. This represents an extension of the study of
\cite{Sijacki2006}, and we refer to this paper for a detailed description of
the models, and the simulation setup, while we here give just a brief
overview.

We consider models of isolated galaxy clusters under a range of different
physical processes. The initial setup consists of a static NFW dark mater halo
\citep{Navarro1996,Navarro1997}, and a gas component which is initially in
hydrostatic equilibrium. The adopted initial gas density profile closely
follows the dark mater profile, except for a slightly softened core. A certain
level of rotation has been included as well, described by a spin parameter of
$\lambda=0.05$. AGN heating has been simulated with a phenomenological
approach in the form of centrally concentrated hot bubbles that are injected
into the ICM in regular time intervals. The basic parameters of the AGN
feedback scenario are the bubble radius, distance from the cluster centre,
duty cycle and bubble energy content. These parameters have been constrained
by recent cluster observations and also by the basic empirical laws of black
hole accretion physics.

We used $10^6$ gas particles to construct initial conditions for a
massive isolated galaxy cluster of mass $10^{15}\,h^{-1} {\rm
M_\odot}$, with a spatial resolution in the gravitational field equal
to $6.5\,h^{-1} {\rm kpc}$. Starting from these identical initial
conditions, we carried out different runs, characterized as follows:
(1) radiative cooling and star formation together with standard
artificial viscosity; (2) cooling, star formation, and AGN-bubble
heating with artificial viscosity; (3) cooling, star formation,
AGN-bubble heating, and physical shear viscosity, based on the
Braginskii parameterization and with a suppression factor that we
varied in the range of $0.3$ to $1.0$\footnote{In these runs, we used
only the physical viscosity, switching off the artificial viscosity
completely. This is here justified because we are simulating an
isolated halo where no strong shocks are present, and due to the fact that
the bubble heating keeps most of the gas above $10^7$K, such that
sufficient shear viscosity is present to evolve the hydrodynamics
correctly, as we explicitly checked.}. The radius of the bubbles was
chosen as $30\,h^{-1} {\rm kpc}$, and they were injected into the ICM
every $10^8\,{\rm yrs}$ along a fixed spatial axis, with an energy
content equal to $2.5 \times 10^{60}\,{\rm ergs}$ per bubble.

\subsection{Radial heating efficiency and profiles}

In Fig.~\ref{profiles_iso}, we show radial profiles of our massive galaxy
cluster after a simulated time of $0.15 \, t_{\rm Hubble}$. Gas density is
plotted in the left panel, mass-weighted temperature in the central panel, and
gas entropy in the right panel. A number of interesting features can be
noticed from the gas profiles. First, regardless of the assumed gas viscosity,
the bubble heating prevents the creation of a strong cooling flow, and gas is
heated efficiently in the inner regions.  Moreover, the spatial extent of the
central region in which AGN feedback alters the gas profiles does not depend
on the level of gas viscosity - in all the runs, bubbles influence the ICM out
to $\sim 150\,h^{-1} {\rm kpc}$. This scale is indicated with a vertical arrow
on the panels. Thus, the radial extent of bubble heating is determined by
other factors, like for example by the initial entropy excess of bubbles with
respect to the surrounding gas, by the injection mechanism, or by the equation
of state of the gas filling the bubbles. Second, it can be seen that an
increase of the gas viscosity produces a systematically stronger heating in
the innermost $\sim 50\,h^{-1} {\rm kpc}$, and this trend is also present at
subsequent simulation output times until $0.25 \, t_{\rm Hubble}$ where we
stopped the simulations. The heating is most prominent in the case of
unsuppressed Braginskii viscosity (orange dot-dashed lines), where an entropy
inversion occurs, and the temperature of the gas keeps increasing until the
very centre. Such a temperature profile is not favoured by observational
findings, suggesting that if the intracluster gas viscosity is indeed so high
than the bubble energy content has to be substantially lower, or the energy
transport from the bubbles into the ICM has to be somehow inhibited, possibly
by magnetic fields.
\begin{figure*}
\centerline{
\vbox{
\psfig{file=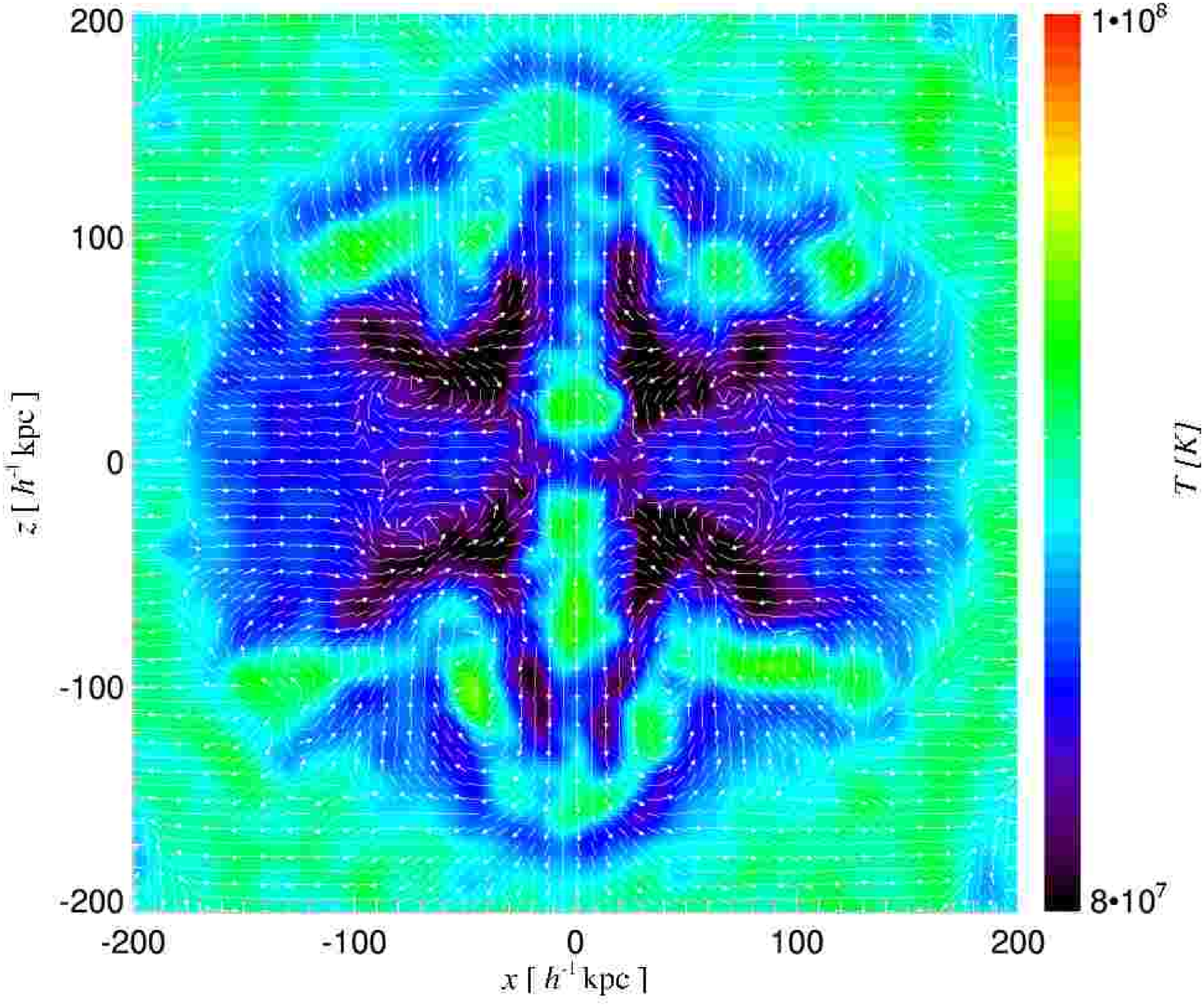,width=12.5truecm,height=11truecm}
\vspace{-0.5truecm}
\psfig{file=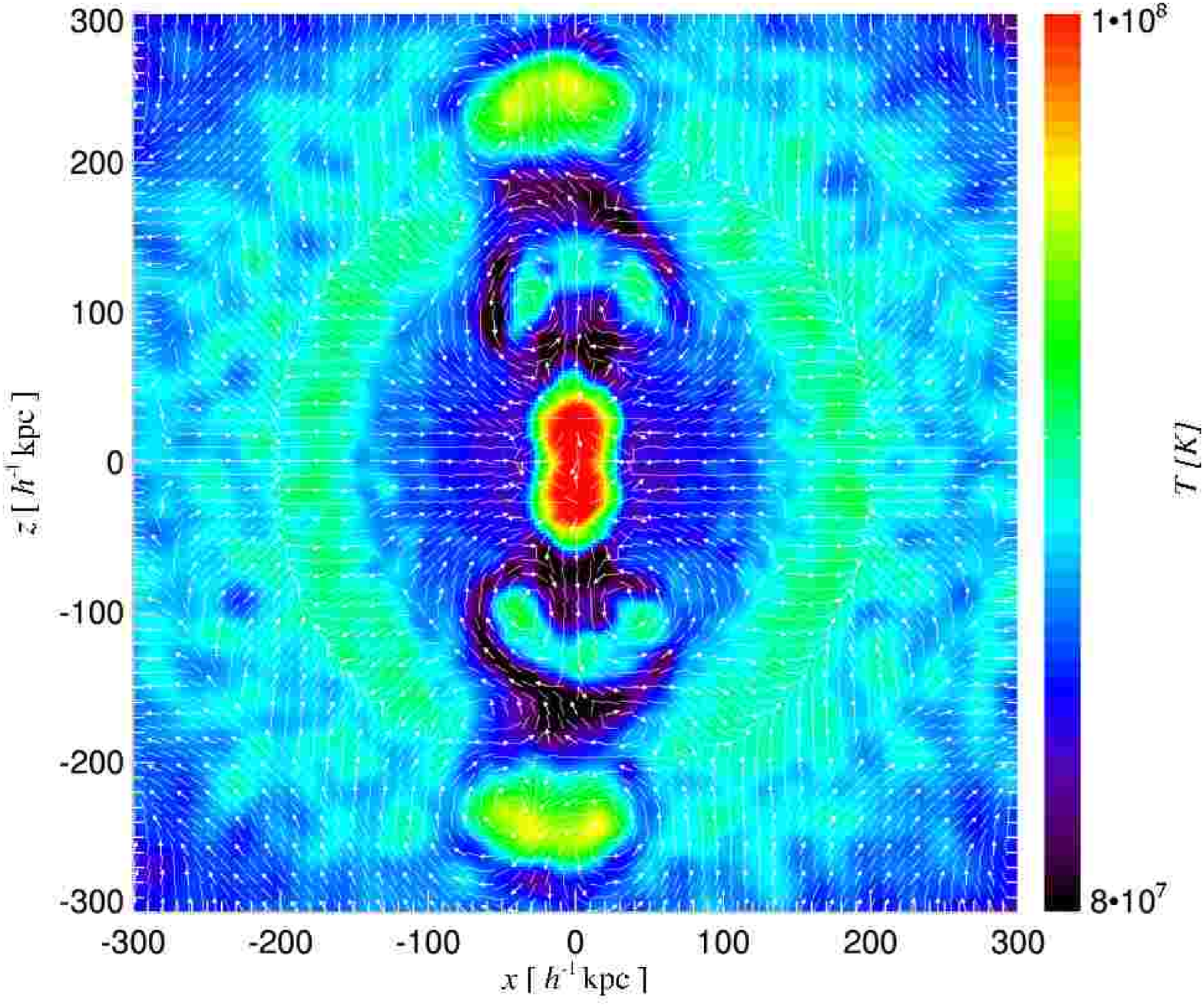,width=12.5truecm,height=11truecm}
}}
\hspace{0.5truecm}
\caption{Mass-weighted temperature maps of a $10^{15}\,h^{-1} {\rm M_\odot}$
  isolated halo, subject to AGN bubble feedback.  The velocity field of the
  gas is over-plotted with white arrows. The maps show how the morphology,
  survival time and maximum distance reached of AGN--driven bubbles depend
  strongly on the amount of physical viscosity assumed: in the upper panel,
  the Braginskii shear viscosity has been suppressed by a factor 0.3, while in
  the lower panel, the simulation has been evolved with the full Branginskii
  viscosity.}
\label{vflow_iso}
\end{figure*}
\begin{figure*}
\centerline{
\vbox{
\hbox{
\psfig{file=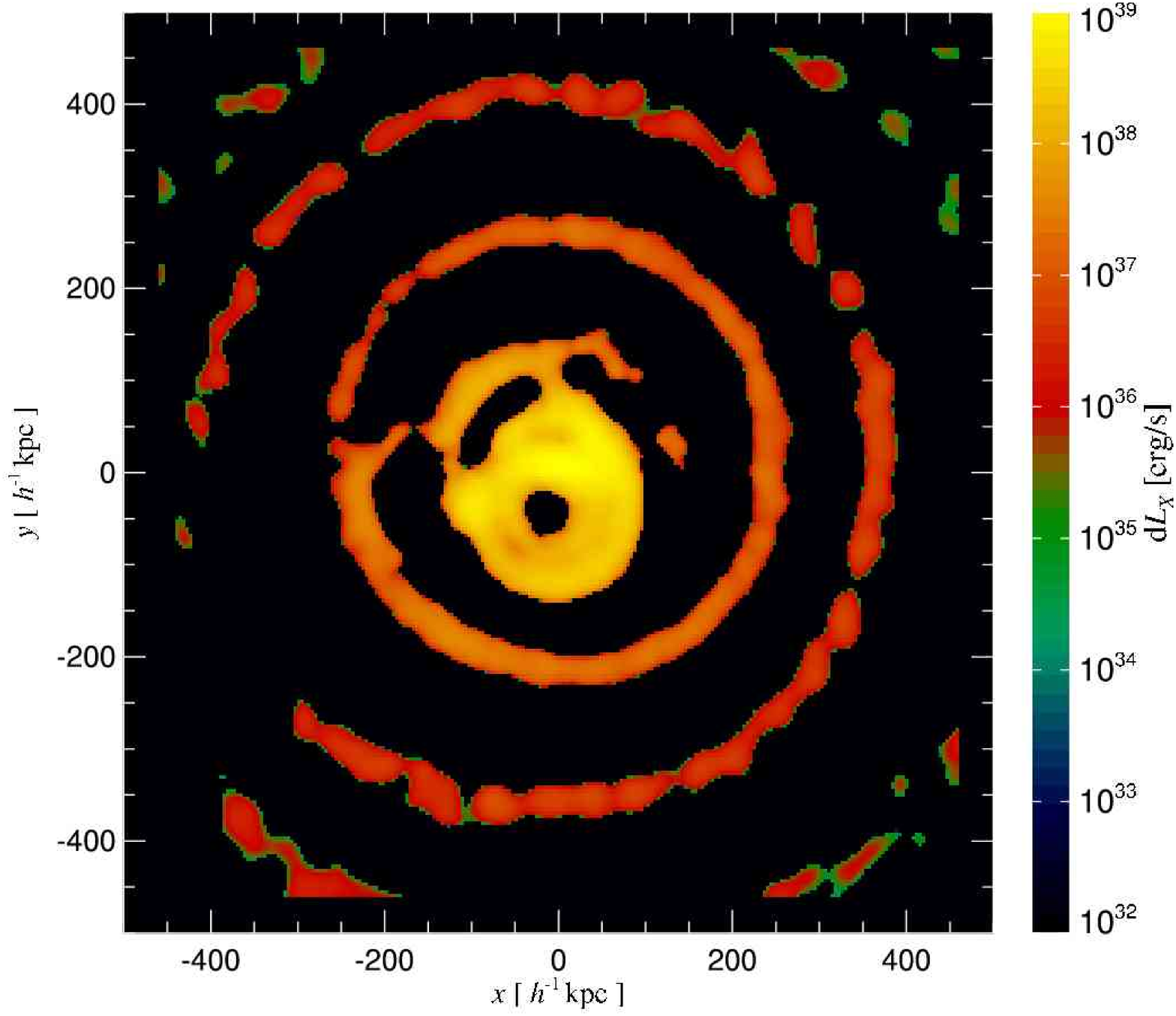,width=8truecm,height=7truecm}
\vspace{-0.5truecm}
\psfig{file=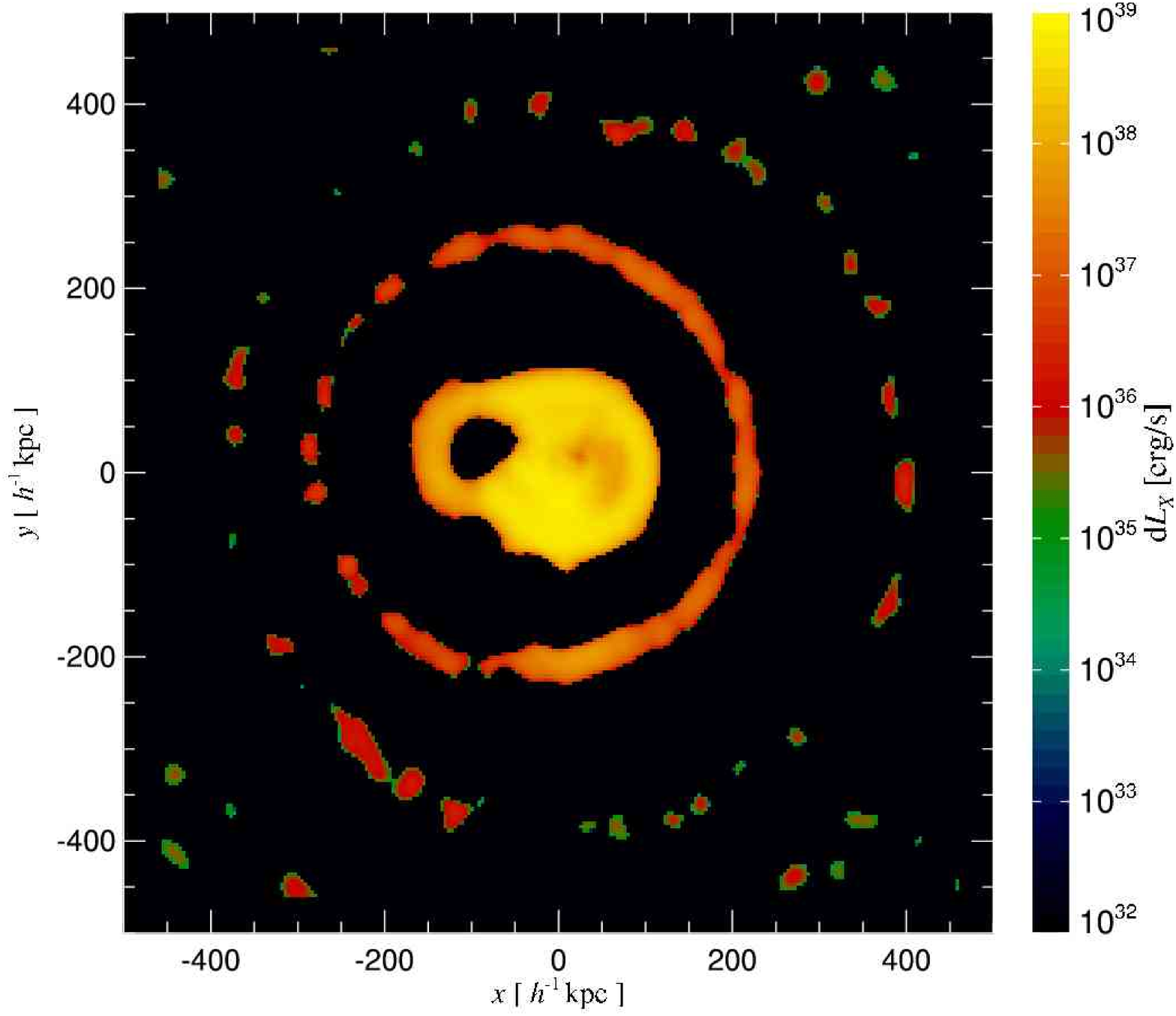,width=8truecm,height=7truecm}
}
\hbox{
\psfig{file=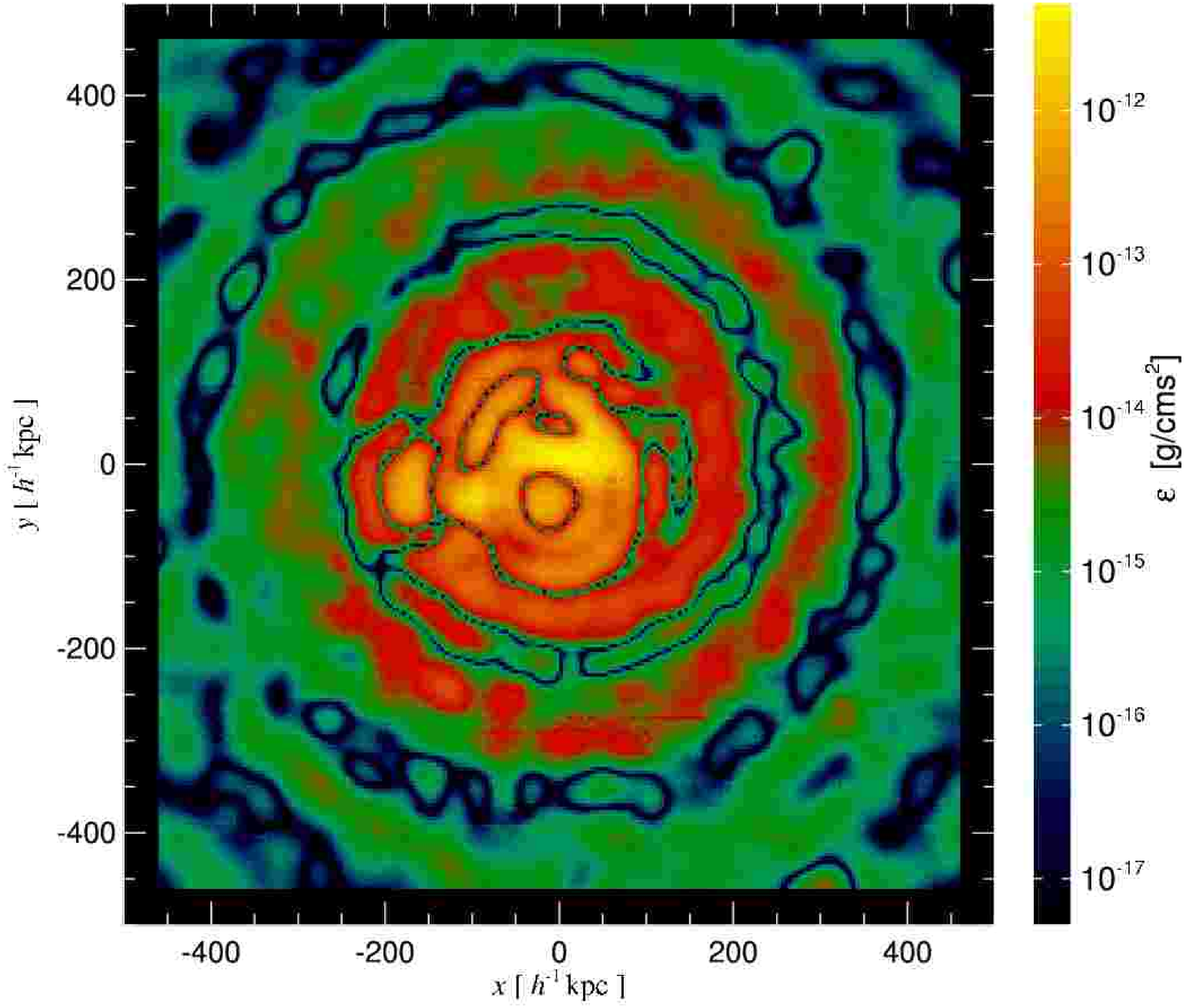,width=8truecm,height=7truecm}
\vspace{-0.5truecm}
\psfig{file=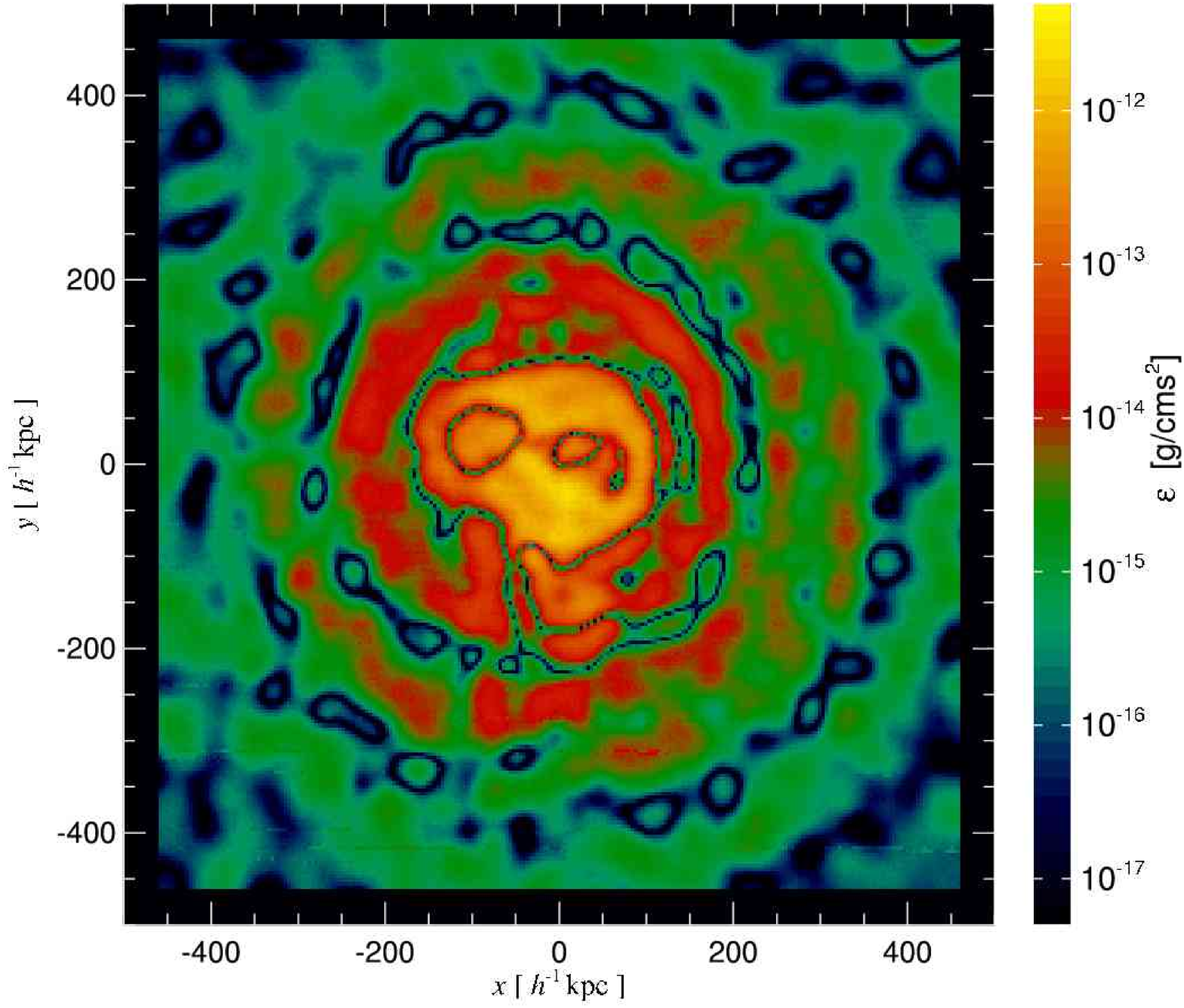,width=8truecm,height=7truecm}
}}}
\hspace{0.3truecm}
\caption{The upper panels show unsharp-masked X--ray emissivity maps of
  a $10^{15}\,h^{-1} {\rm M_\odot}$ isolated cluster with AGN bubble
  feedback. On the left, a case with lower viscosity is shown (suppression
  factor $f = 0.3$), while on the right, a case with the maximum value of the
  shear viscosity is displayed (suppression factor $f = 1.0$).  The sound
  waves generated by the bubbles in the ICM are clearly more efficiently
  dissipated when the ICM has a higher viscosity. The panels on the bottom
  illustrate the projected energy density in sound waves for the same cases,
  estimated as explained in the text. On these maps, both compressed and
  rarefied regions are visible, containing a significant amount of acoustic
  energy density. Nevertheless, the total energy of the sound waves is not
  substantial in our models, and amounts to only a small fraction of the
  thermal energy of an injected bubble.}
\label{soundwaves}
\end{figure*}

Another interesting feature of bubble heating in a viscous ICM can be noticed
when the bubble morphologies and the radial heating efficiency are examined in
more detail. In Fig.~\ref{vflow_iso}, we show mass-weighted temperature maps
of the central cluster regions, for the case of Braginskii viscosity with
suppression factor of $0.3$ (upper panel) and for unsuppressed Braginskii
shear viscosity (lower panel). The velocity flow pattern is indicated with
white arrows on these maps. Even though the radial extent of the bubble
heating is similar for different magnitudes of internal friction, the
morphologies of evolved bubbles, their maximum clustercentric distance reached
and their survival times vary. When the gas viscosity is as high as the full
Braginskii value, the bubbles rise up to $\sim 300\,h^{-1} {\rm kpc}$ in the
cluster atmosphere without being disrupted, and traces of two up to three past
bubble episodes can be detected, indicating that the bubbles survive at least
as long as $\sim 2 \times 10^8\,{\rm yrs}$.  However, when the gas viscosity
is lowered (upper panel of Fig.~\ref{vflow_iso}) bubbles typically start to
disintegrate at $\sim 150\,h^{-1} {\rm kpc}$ and multiple bubble events can
typically not be identified. 

This suggests that a relatively high amount of gas viscosity may be needed to
explain the recent observations of the Perseus cluster
\citep[e.g.][]{Fabian2006}, where several bubble occurrences have been
detected.  However, an alternative explanation could be that the bubbles are
stabilized against fluid instabilities by magnetic fields at their
interface with the ICM instead of by viscosity. A relativistic particle
component (cosmic rays) filling the bubble will also change the dynamical
picture. Nevertheless, it is interesting that the observed morphology of
bubbles can in principle constrain the level of ICM viscosity, an aspect that
we plan to explore further in a future study.

In the velocity fields shown in Fig.~\ref{vflow_iso}, it can be seen that the
flow of the gas in the wake of the bubbles is approximately laminar, while at
the bubble edges the velocity field shows a significant curl component.  This
perturbed motion is not only present for the most recently injected bubbles,
but also for the bubbles that are already $ \sim 2 \times 10^8\,{\rm yrs}$
old, albeit with a smaller magnitude. In the case of the full Braginskii
viscosity the magnitude of the velocity perturbations induced by the bubbles
is of the order of $ \lsim\, 100\,{\rm km\,s^{-1}}$ for the recent bubbles,
while it decreases to $ \lsim\, 20 \,{\rm km\,s^{-1}}$ for the older ones.

\subsection{Sound waves dissipation}

We also examined how the occurrence of sound waves produced by the bubbles and
the associated non-local heating is influenced by different amounts of ICM
viscosity.  In Fig.~\ref{soundwaves}, we show unsharp-masked images of the
X--ray emissivity, produced by subtracting a map smoothed on a $50\,h^{-1}
{\rm kpc}$ scale from the original luminosity map. It is clear that for higher
gas viscosity (right panel, unsuppressed Braginskii value) the damping of
sound waves in the central region is stronger than for a simulation with lower
viscosity (left panel: $0.3$ of Braginskii viscosity). Nevertheless, the
radial profiles of the gas entropy show that only in the inner $\sim
50\,h^{-1} {\rm kpc}$ a more efficient heating of the ICM can be observed when
the viscosity is increased. This suggests that the energy content of the sound
waves produced by the bubbles is not very large and probably not capable of
providing significant heating at larger radii.

\begin{table*}
\bc
\begin{tabular}{crrccccc}
\hline \hline Simulation & $N_{\rm HR}$ & $N_{\rm gas}$ & $m_{\rm DM}$
[$\,h^{-1}{\rm M}_\odot\,$] & $m_{\rm gas}$ [$\,h^{-1}{\rm
M}_\odot\,$] & $z_{\rm ini}$ & $z_{\rm fin}$ & $\epsilon$
[$\,h^{-1}{\rm kpc}\,$]\\ \hline g1/g8 & $4937886$ & $4937886$ &
$1.13\times 10^9$ & $0.17\times 10^9$ & $60$ & $0$ & $5.0$ \\ \hline
\hline
\end{tabular}
\caption{Numerical parameters of the cosmological galaxy cluster
  simulations used in this study. The values listed from the second to
  the fifth column refer to the number and to the mass of high
  resolution dark matter particles and of gas particles. Note that the
  actual values of $N_{\rm gas}$ and $m_{\rm gas}$ may vary in time
  due to star formation, if present.  The last three columns give the
  initial and final redshifts of the runs, and the gravitational
  softening length $\epsilon$.}
\label{tab_simpar}
\ec
\end{table*}

In order to put more stringent constraints on the influence of the sound
waves, we have estimated their energy content by evaluating \citep{LandauFM}
\be \int E_s \,{\rm d}V \, = \, \int \rho_0 v^2 \,{\rm d}V \, = \, \int
\frac{{\rho'}^2}{\rho_0} c_s^2 \,{\rm d}V, \ee where $E_s$ is the sound energy
density, $\rho_0$ is the unperturbed gas density, $v$ is the fluid velocity,
$\rho'$ is the change in gas density due to the sound waves, and $c_s$ is the
sound speed. Strictly speaking, this equation is valid only for travelling
plane waves, but it should still provide us with a reliable order of magnitude
estimate of the sound waves energy in our geometrically more complex case. We
computed projected energy density maps (see lower panels of
Fig.~\ref{soundwaves}) by taking the smoothed density field for $\rho_0$,
while we estimated $\rho'$ as the difference between the actual density field
and the smoothed one. Both compressed regions (which correspond to the rings
in the upper panels) and rarefied regions store a considerable amount of
energy. However, it should be noted that the sound wave energy in the
rarefaction regions is overestimated when computed in this way, because the
bubbles themselves contribute to it, being underdense with respect to the
background and having similar dimension to the smoothing scale. Nevertheless,
when the total sound wave energy is estimated in this way we obtain $\sim 5
\times 10^{59}\,{\rm ergs}$, which is only a small fraction of the initially
injected bubble energy.  As pointed out by \cite{Churazov2002}, a number of
the same order of magnitude is obtained if as a crude estimate of the sound
wave energy one considers the sound waves generated by the motion of a
solid sphere through a medium of a given density $\rho$ \citep{LandauFM}. 

Based on these findings, the viscous damping of sound waves provides
only an insignificant contribution in coupling the AGN-injected energy
into the ICM.  Note that in these models of isolated halos we have
deliberately included physical shear viscosity only. Thus, our
estimate for the damping of sound waves is not affected by any
residual artificial viscosity.  A caveat, however, is that our
simulations do not self-consistently model the initial phase of bubble
injection, where the AGN-jet deposits its energy and inflates the
bubble.  It is conceivable that the associated processes produce
energetically more important sound waves and weak shocks, which could
then increase the importance of viscous damping of sound waves
compared to the result found here.

\section{Cosmological simulations of viscous galaxy clusters} \label{Cosmological}

In this section we discuss the effects of internal friction on clusters of
galaxies formed in fully self-consistent cosmological simulations. We have
carried out a variety of runs that follow different physical processes,
including non--radiative hydrodynamical simulations, described in detail in
Section~\ref{Non--radiative}, and runs with radiative gas cooling, star
formation and feedback processes, which are discussed in
Section~\ref{Cooling}. For all these simulations, we carried out matching
pairs of runs without physical viscosity and with physical shear viscosity
(using the Braginskii parameterization with a suppression factor of 0.3), in
order to be able to clearly identify differences due to the viscous
dissipation processes.

For our simulations, we selected two massive galaxy clusters that have been
extracted from a cosmological $\Lambda$CDM model with a boxsize of
$479h^{-1}{\rm Mpc}$ \citep{Yoshida2001,Jenkins2001}, and were prepared by
\citet{Dolag2004} for resimulation at higher resolution using the Zoomed
Initial Condition technique \citep{Tormen1997}. In Tables~\ref{tab_simpar}
and~\ref{tab_clusterpar}, we summarize the basic parameters of the simulations
and the main physical properties of the galaxy clusters. The cosmological
parameters of the simulations correspond to a concordance $\Lambda$CDM model
with $\Omega_m=0.3$, $\Omega_b=0.04$, $\sigma_8=0.9$, and $H_0=70 \, {\rm
  km\,s}^{-1}{\rm Mpc}^{-1}$ at the present epoch.

\begin{table}
\bc
\begin{tabular}{lccccc}
\hline
\hline
Cluster & $R_{\rm 200}$ & $M_{\rm 200}$ & $T_{\rm mw}$
& $L_{\rm X}$  \\
& [$\,h^{-1}{\rm kpc}\,$] & [$\,h^{-1}{\rm M}_\odot\,$] & [K] &
 [$\,\rm ergs^{-1}\,$] \\
\hline
g1\_csf  & $2857$ & $1.63\times10^{15}$ & $7.3\times10^7$ &
$1.0\times10^{45}$\\
g1\_csfv  & $2832$ & $1.58\times10^{15}$ & $8.1\times10^7$ &
 $1.0\times10^{45}$\\
g8\_ad  & $3306$ & $2.52\times10^{15}$ & $9.7\times10^7$ &
 $1.1\times10^{46}$\\
g8\_adv  & $3276$ & $2.45\times10^{15}$ & $1.1\times10^8$ &
 $4.5\times10^{45}$\\
\hline
\hline
\end{tabular}
\caption{Physical properties of our sample of simulated galaxy
  clusters at $z=0$ and at $200\rho_c$. For two different galaxy
  clusters, labelled in the first column, and for the different runs,
  cluster radius, total mass, mass--weighted gas temperature and
  X--ray luminosity are listed, respectively. Subscripts in the first
  column denote runs including different physics, namely cooling and
  star formation for the g1 cluster, and non--radiative gas
  hydrodynamics for the g8 cluster, in both cases also with Braginskii
  shear viscosity with suppression factor of 0.3.}
\label{tab_clusterpar}
\ec
\end{table}

\subsection{Non--radiative simulations} \label{Non--radiative}

\begin{figure*}
\bc
\centerline{\includegraphics[width=16truecm,height=20truecm]{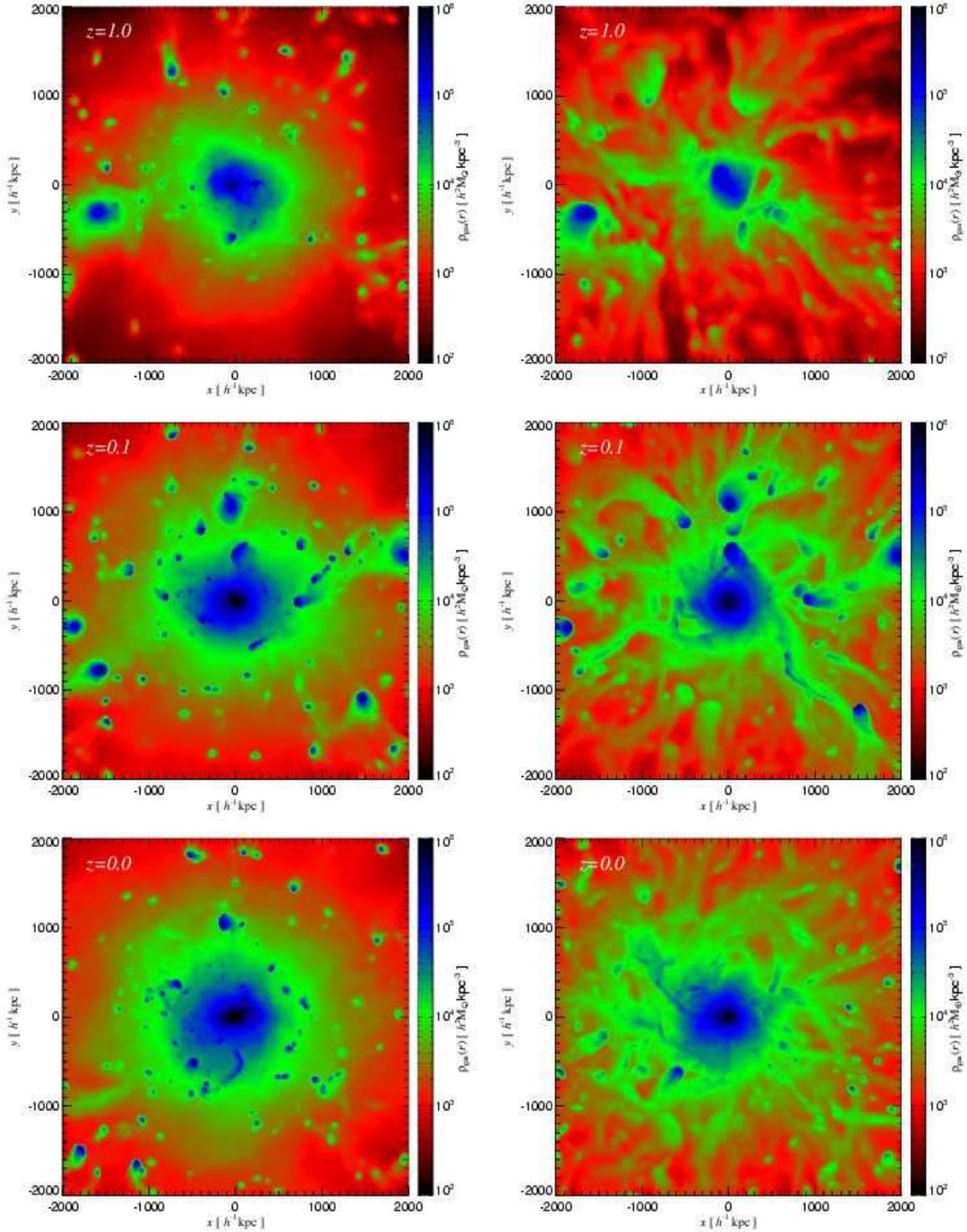}}
\caption{Projected gas density maps of the g8 galaxy cluster
  simulation at redshifts $z=1.0$, $z=0.1$ and $z=0.0$, as indicated in the
  upper-left corner of the panels. The panels in the left column show the gas
  density distribution in the case of a non--radiative run, while the panels
  in the right column give the gas density distribution when Braginskii shear
  viscosity is ``switched-on'', using a suppression factor of 0.3. It is
  evident from these panels that the presence of a modest amount of shear
  viscosity has a significant impact on the gas distribution, removing more
  gas from infalling structures when they enter the massive halo, and producing
  pronounced gaseous tails behind them.}
\label{g8_rhomaps}
\ec
\end{figure*}

In Fig.~\ref{g8_rhomaps}, we show projected gas density maps at different
redshifts for our non-radiative cluster simulations. It is evident that
already at early times, at around $z \sim 5$, the gas distribution in the
presence of shear viscosity (panels on the right) starts to deviate
substantially from the corresponding simulation without internal friction
(panels on the left). Also, the amount of gas that is bound to dark matter
subhalos is reduced, and there appears to be more diffuse gas in the outskirts
of massive objects. Furthermore, small structures that are falling into the
most massive halo at each epoch (located in the centre of the panels), lose
their gas content more quickly due to the shear forces, and feature prominent
tails that extend up to several hundred kiloparsec. These general
features are present
at all epochs from $z \sim 5$ to $z = 0$.  At low redshifts, however, the
central parts of the main halo appear quite similar, although the central
density is somewhat reduced in the simulation with physical viscosity, while
there appears to be more diffuse gas with a smaller number of prominent gas
concentrations in the outskirts. These trends can be readily understood as a
consequence of viscous dissipation, which increases the stripping of gas and
helps to expel gas from shallow dark matter potential wells in the infall
regions of larger structures.

\begin{figure*}
\centerline{
\vbox{
\hbox{
\psfig{file=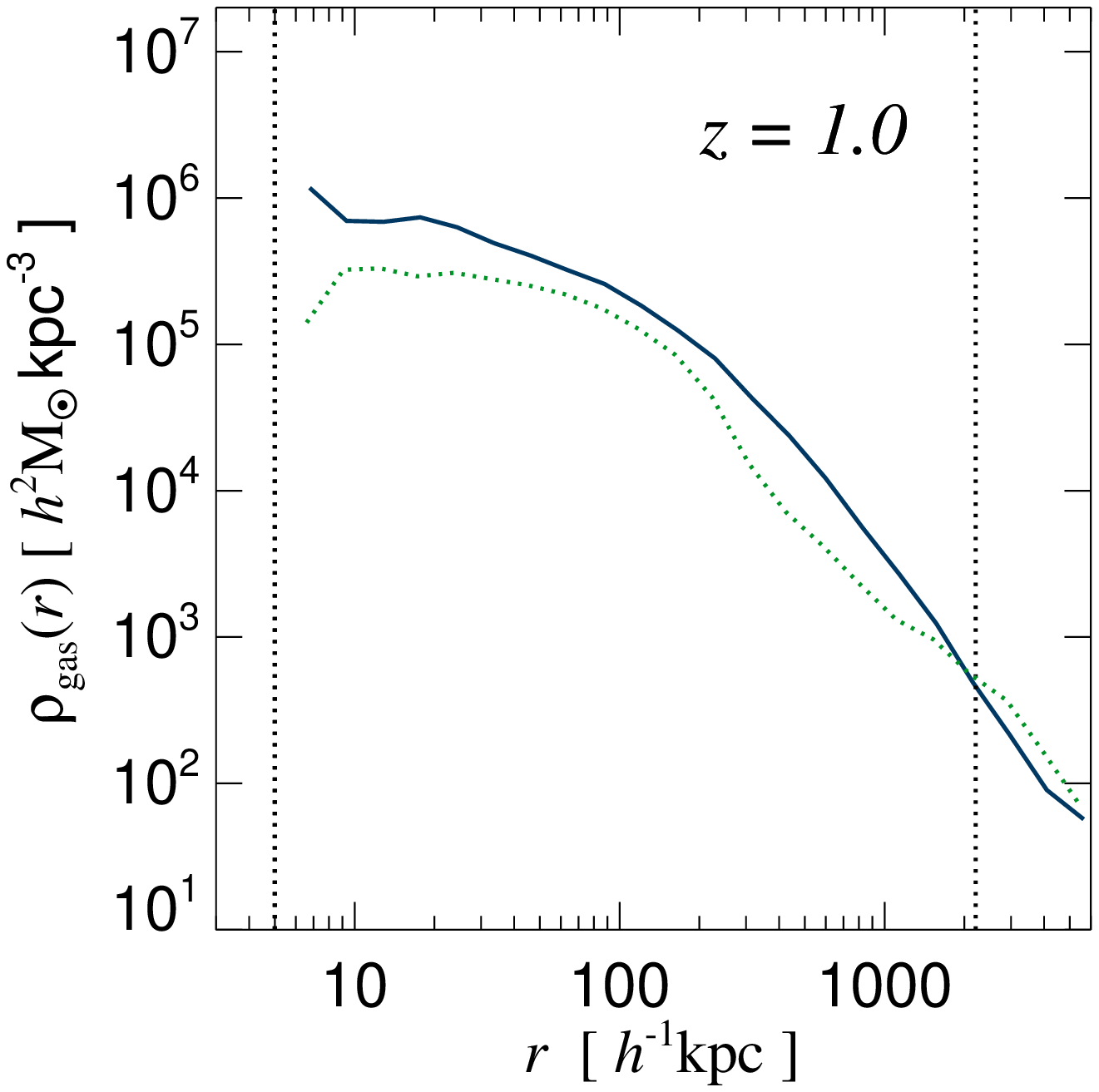,width=7.5truecm,height=7.5truecm}
\vspace{-0.5truecm}
\psfig{file=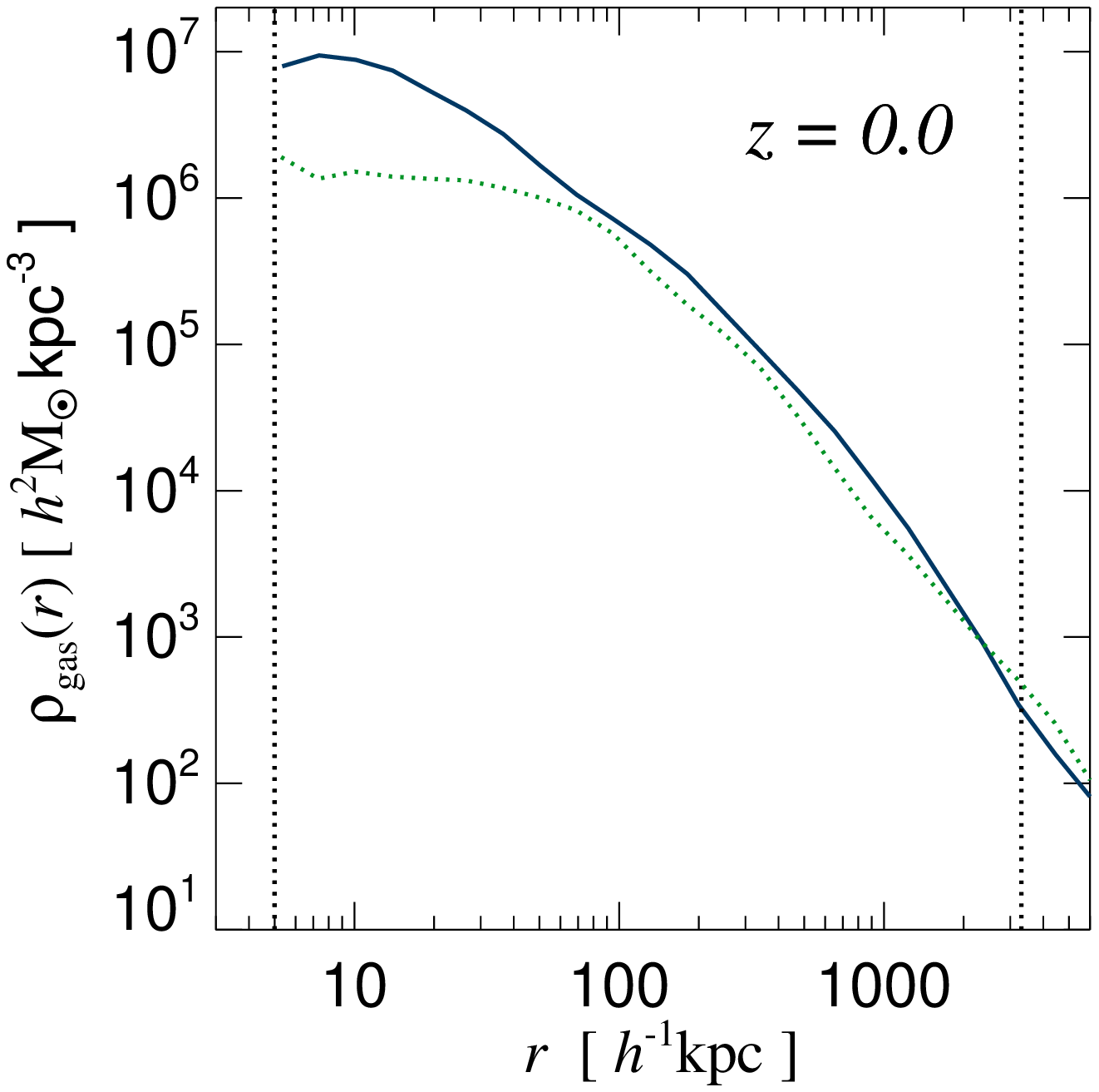,width=7.5truecm,height=7.5truecm}
}
\hbox{
\psfig{file=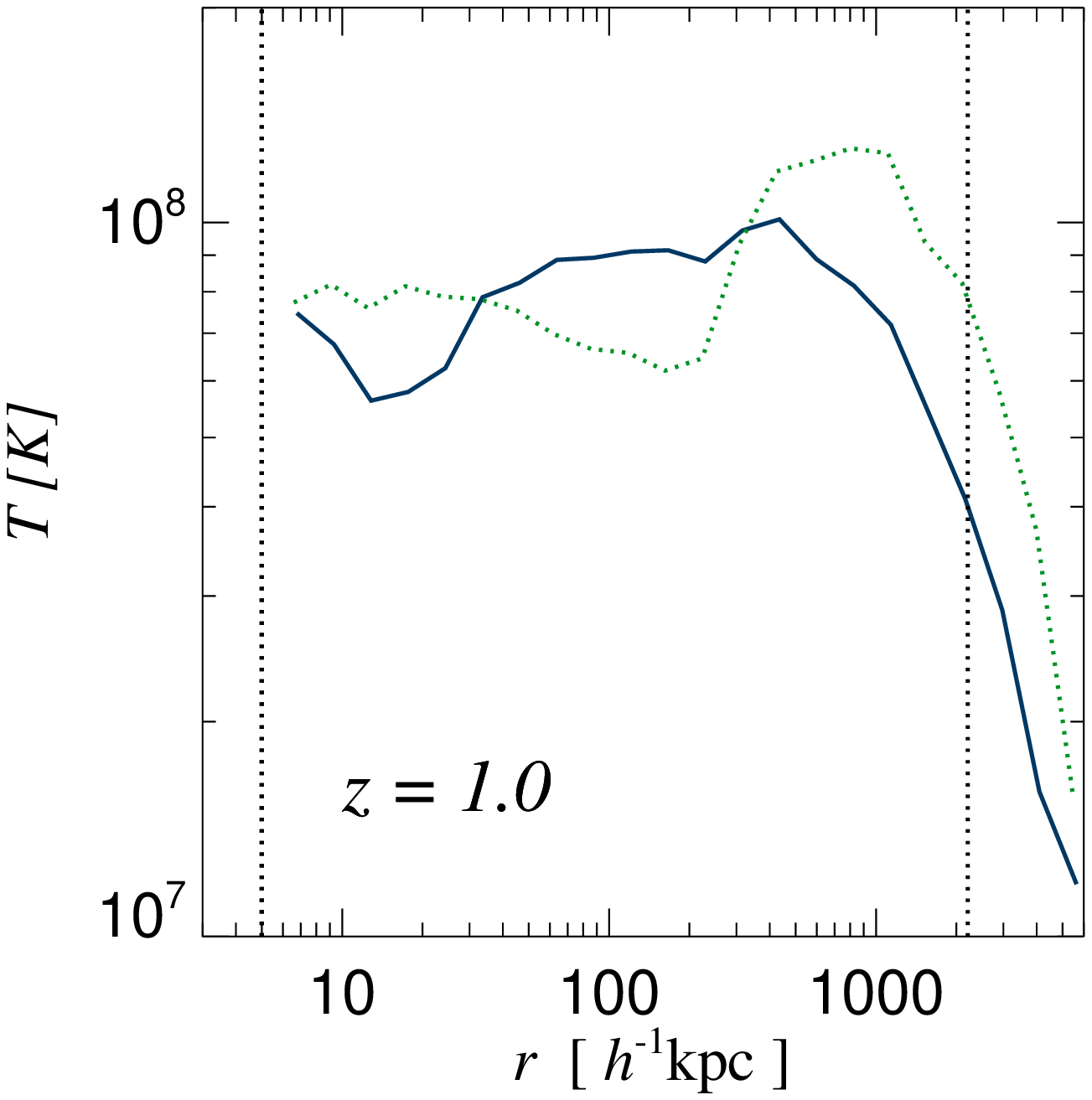,width=7.5truecm,height=7.5truecm}
\vspace{-0.5truecm}
\psfig{file=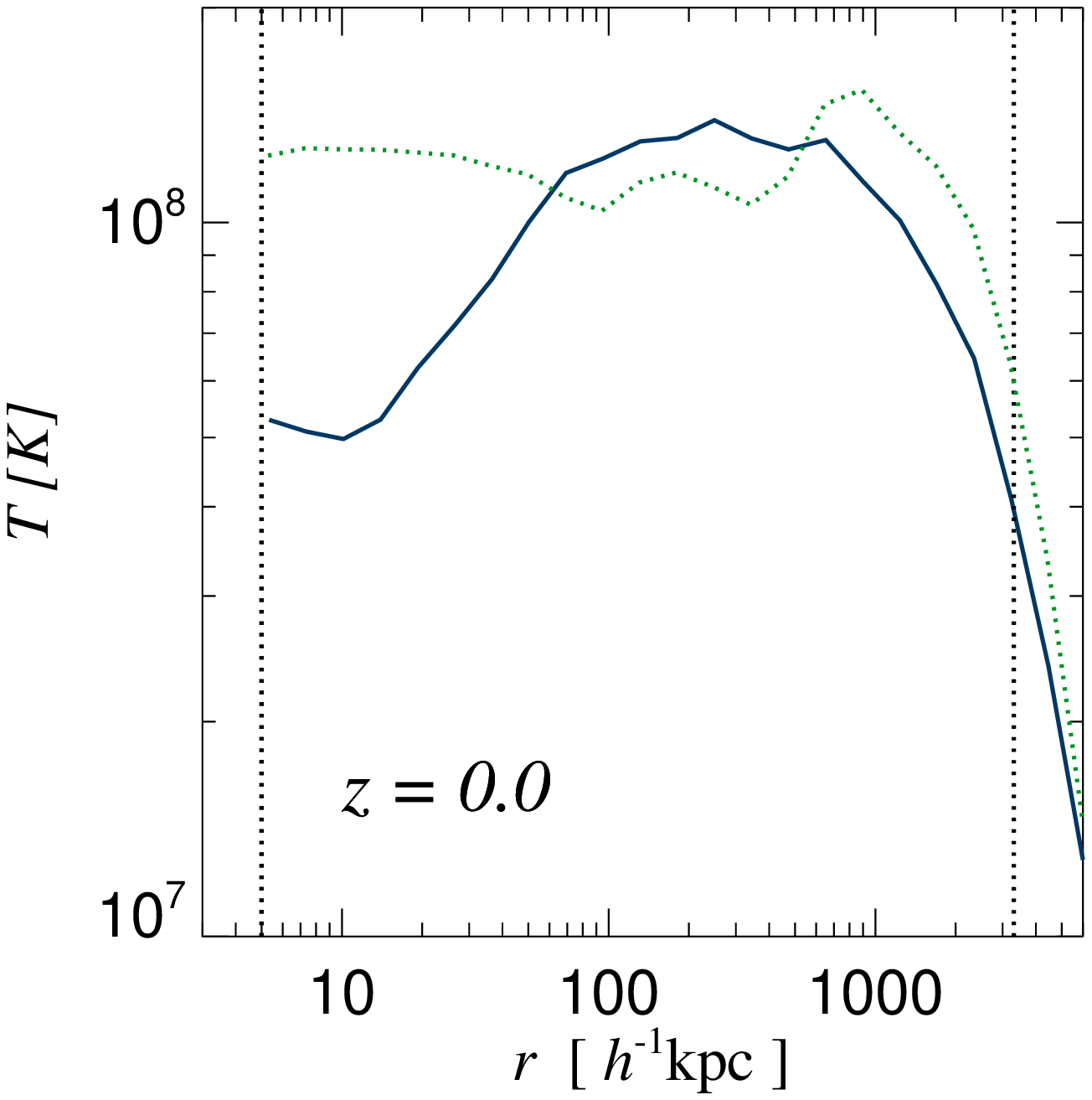,width=7.5truecm,height=7.5truecm}
}
\hbox{
\psfig{file=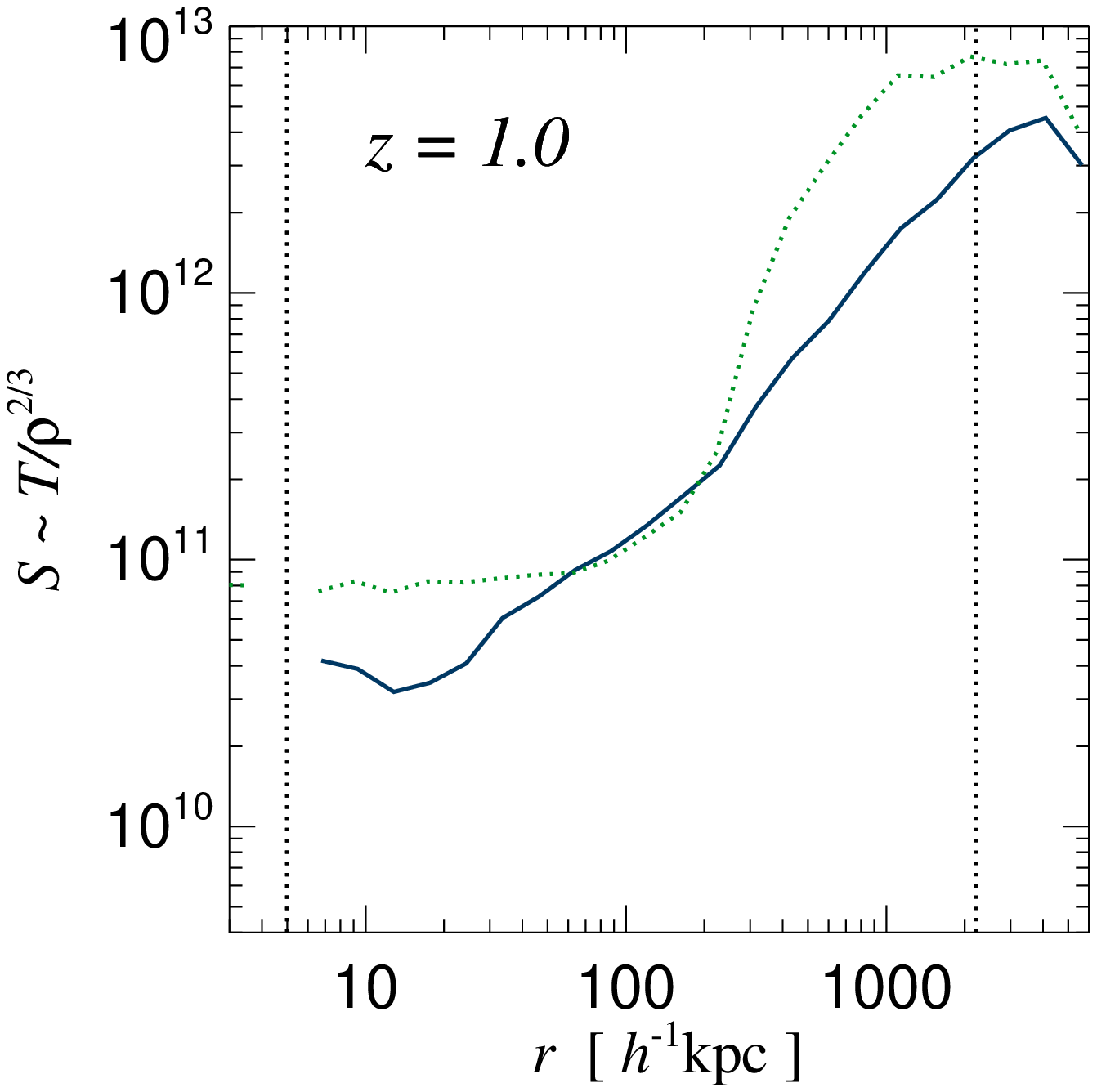,width=7.5truecm,height=7.5truecm}
\vspace{-0.5truecm}
\psfig{file=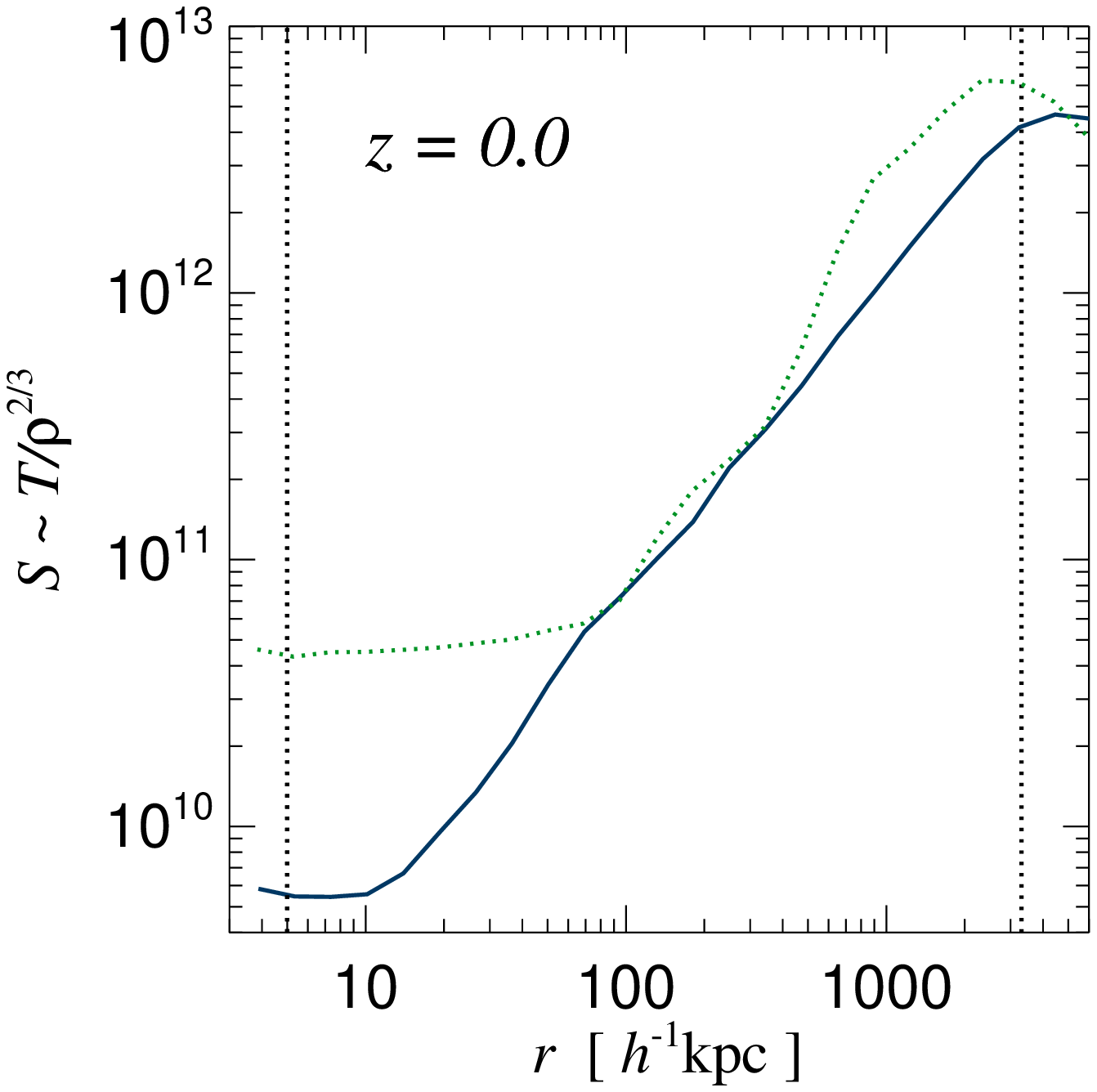,width=7.5truecm,height=7.5truecm}
}
}}
\hspace{1.5truecm}
\caption{Radial profiles of gas density, mass-weighted temperature, and
  entropy of the `g8' galaxy cluster simulation. The blue continuous lines are
  for the non--radiative run, while the dotted green lines are for the run
  with additional Braginskii shear viscosity with suppression factor of 0.3.
  The panels in the left column refer to the gas profiles evaluated at
  $z=1.0$, while the ones in the right column are for $z=0$. The dotted
  vertical lines denote the gravitational softening length and the virial
  radius at the given redshift.}
\label{g8_profiles}
\end{figure*}

A more quantitative analysis of how viscosity affects the thermodynamics of
clusters is obtained by studying the radial profiles of gas density,
mass-weighed temperature, and entropy.  In Fig.~\ref{g8_profiles}, we compare
these profiles at two different epochs, $z=1$ and $z=0$.  The solid blue line
refers to the non-radiative run, while the dotted green line gives the result
when physical shear viscosity is included. The effects of viscosity increase
systematically with time, and manifest themselves in a reduction of the gas
density throughout the cluster.  The suppression is particularly strong in the
very centre, where it reaches almost an order of magnitude at $z=0$. At early
times, the temperature is mainly boosted in the outer regions of the cluster,
while for $z < 0.5$, also the central gas temperature starts to be
significantly increased. As a consequence, the entropy profile shows two
different features: At early times, the central rise of entropy is caused by a
lower gas density, while at late times, the entropy is even more enhanced in
the inner regions, out to a radius of $\sim 100 \,h^{-1}{\rm kpc}$, because of
a lower gas density and an increased temperature. In the outer parts of the
halo, the increase of the gas entropy is always a result of the joint action
of temperature and density change.  We will come back to a discussion of the
physical reason for this behaviour in Section~\ref{Cooling}.

As a consequence of the strong modification of the density and temperature
structure of the cluster in the viscous case, we also find a significant
reduction of the X--ray luminosity. This is  reflected in
Table~\ref{tab_clusterpar}, which lists some of the basic properties of these
clusters. Interestingly, a similar change of the X--ray luminosity is not
found in the case of the simulations that also include cooling and star
formation, which we will discuss next.

\subsection{Simulations with cooling and star formation} \label{Cooling}

\begin{figure*}
\centerline{
\vbox{
\hbox{
\psfig{file=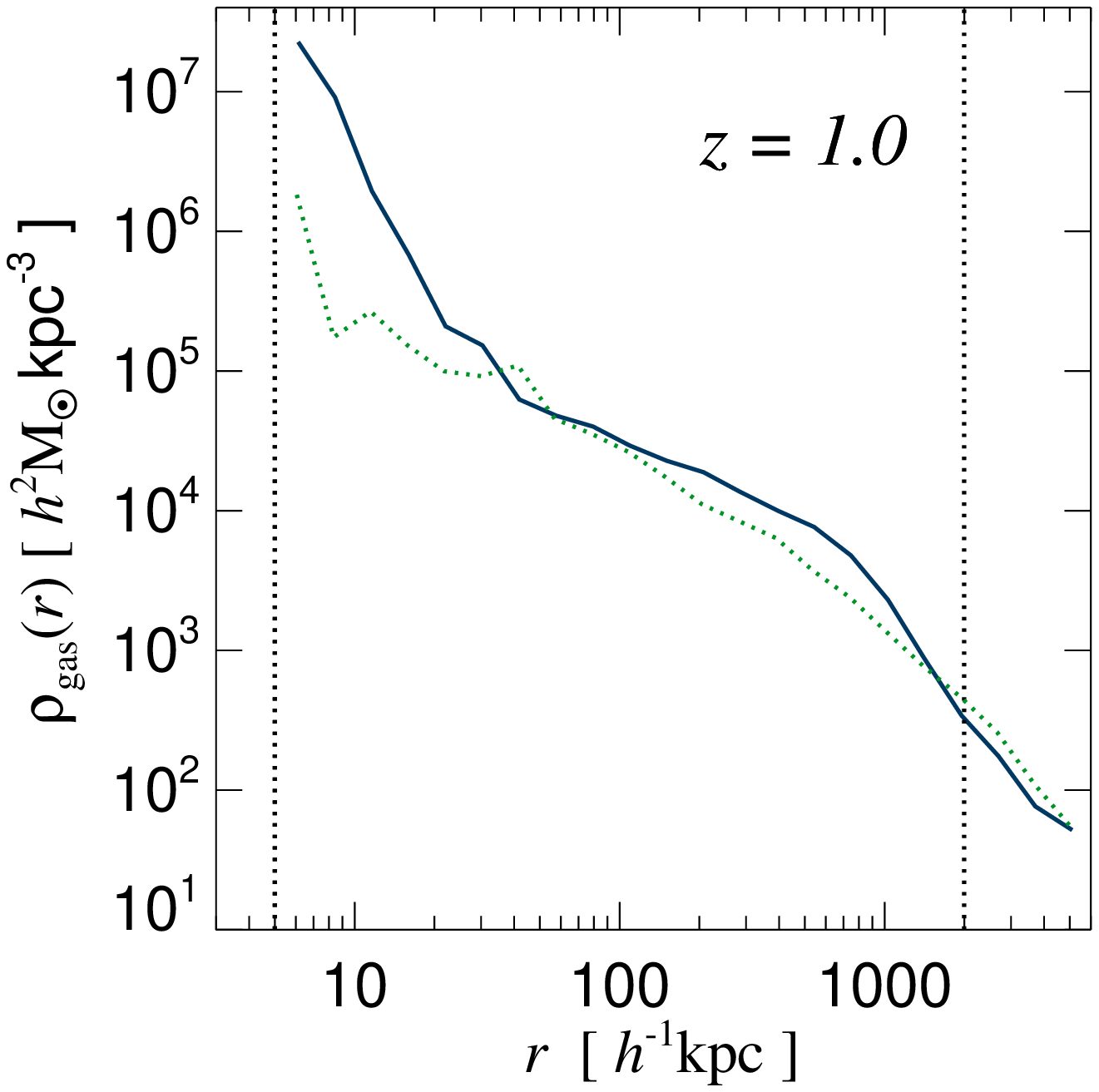,width=7.7truecm,height=7.7truecm}
\vspace{-0.5truecm}
\psfig{file=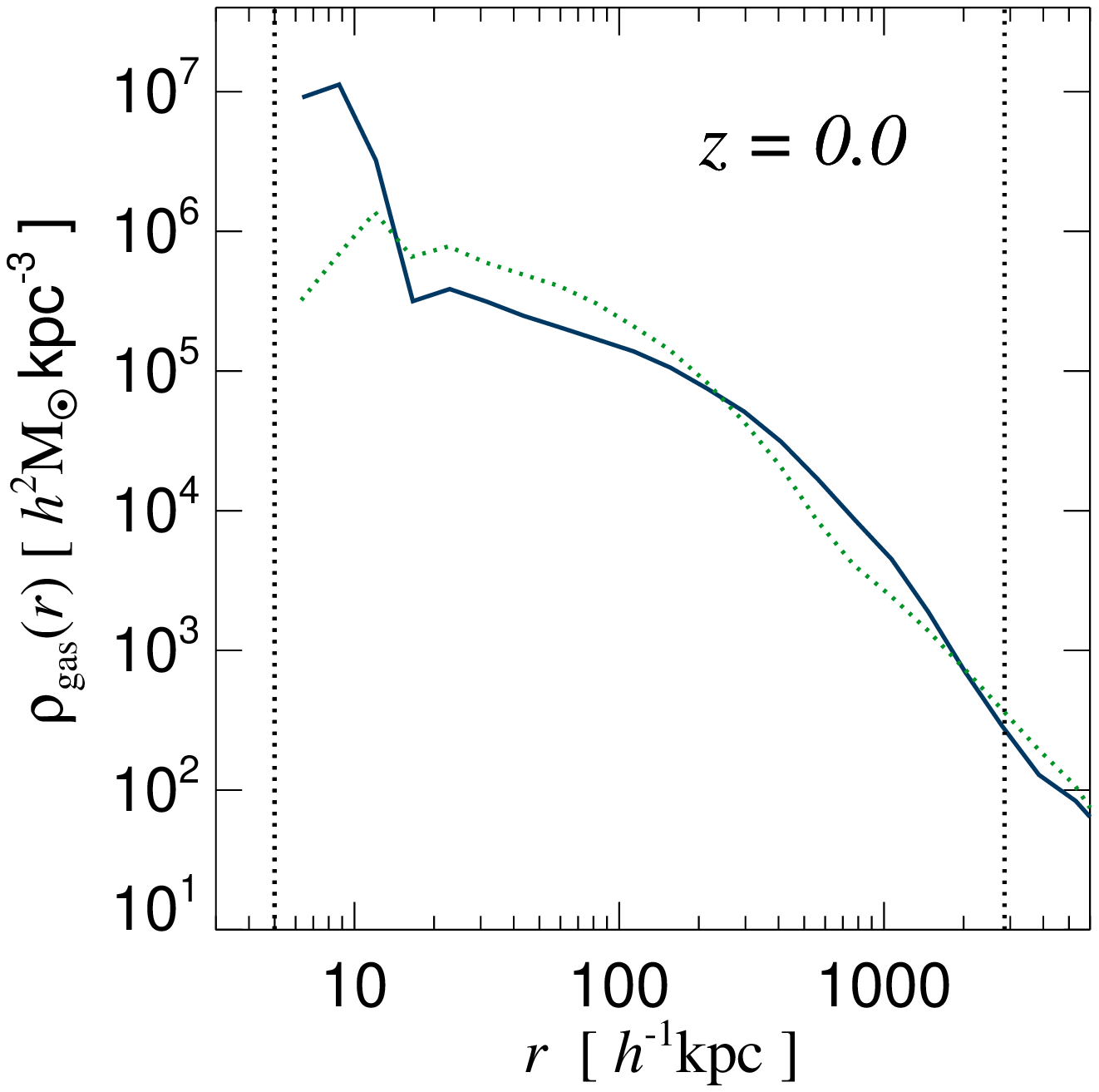,width=7.7truecm,height=7.7truecm}
}
\hbox{
\psfig{file=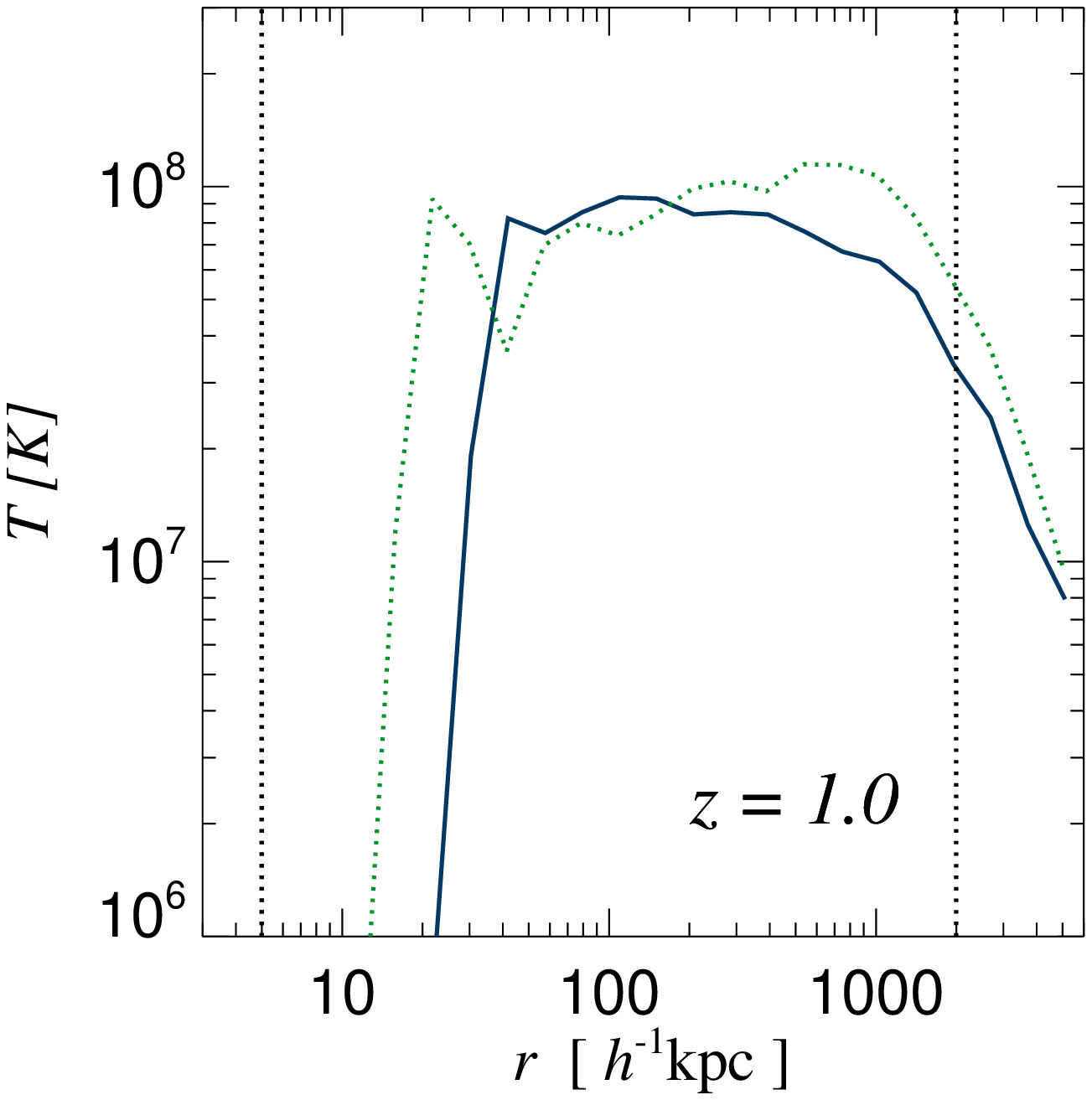,width=7.7truecm,height=7.7truecm}
\vspace{-0.5truecm}
\psfig{file=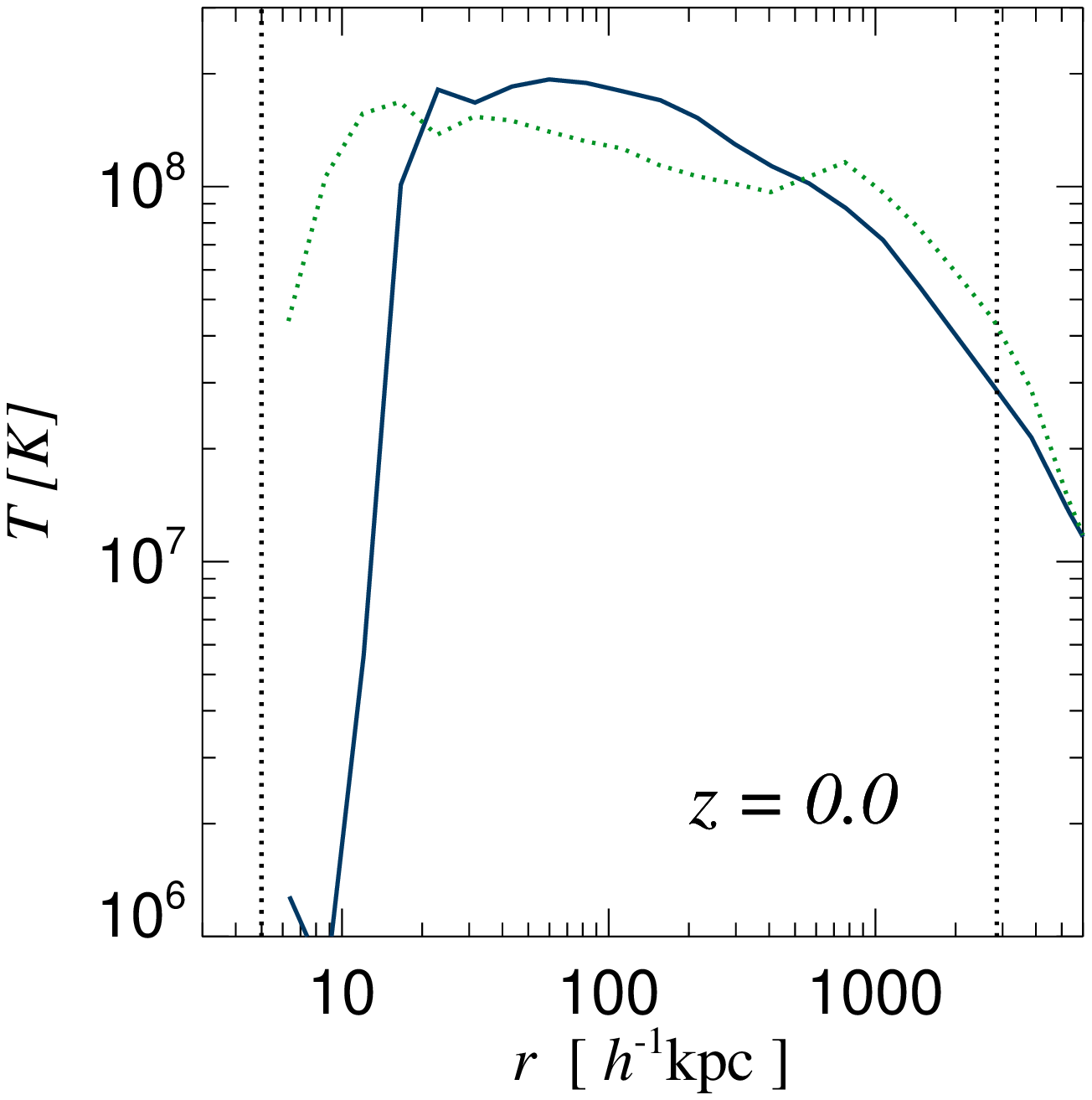,width=7.7truecm,height=7.7truecm}
}
\hbox{
\psfig{file=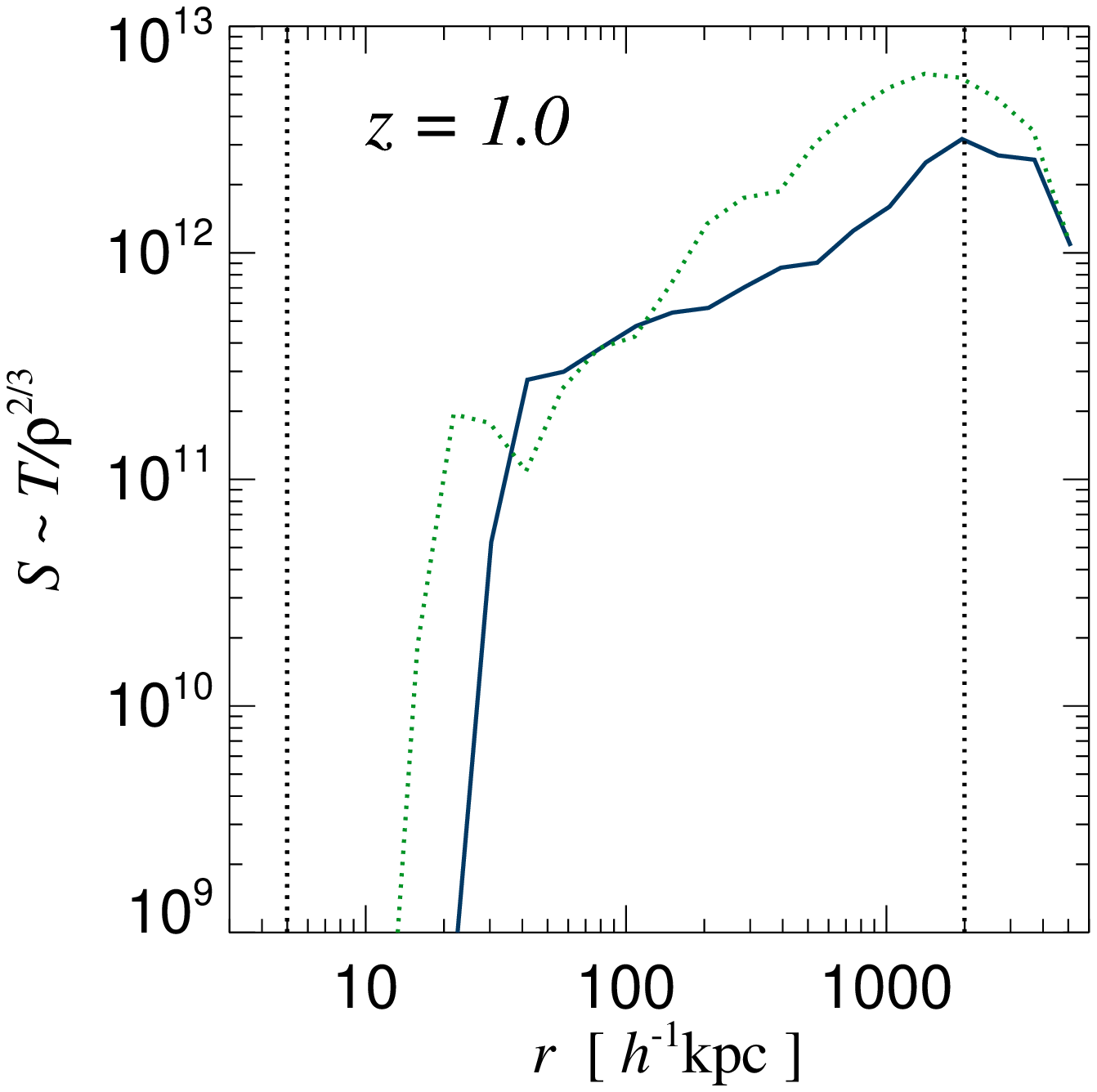,width=7.7truecm,height=7.7truecm}
\vspace{-0.5truecm}
\psfig{file=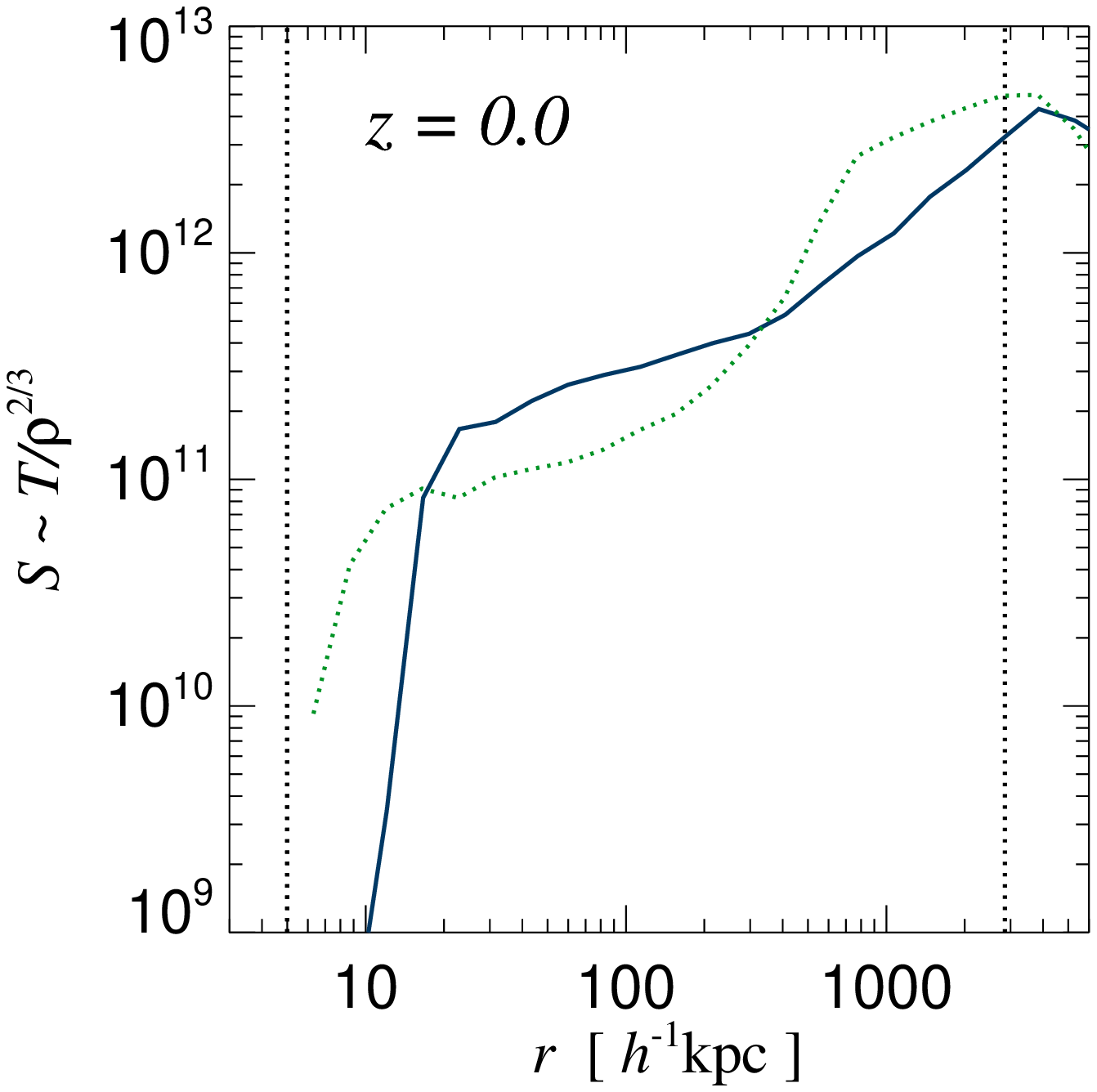,width=7.7truecm,height=7.7truecm}
}
}}
\caption{Radial profiles of gas density, mass-weighted temperature,
  and entropy of the `g1' galaxy cluster simulation. The blue continuous lines
  are for the run with cooling and star formation, while the dotted green
  lines are for the run with additional Braginskii shear viscosity with
  suppression factor of 0.3. The panels in the left column refer to the gas
  profiles evaluated at $z=1.0$, while the panels in the right column are for
  $z=0$. The dotted vertical lines indicate the gravitational softening length
  and the virial radius at the given redshift. It can been seen that the
  viscous effects are similar to the non--radiative case at high redshifts,
  while they differ significantly at low $z$ due to the formation of a central
  cooling flow.}
\label{g1_profiles}
\end{figure*}

In Fig.~\ref{g1_profiles}, we show the radial gas profiles of basic
thermodynamic quantities of our cluster simulations that included radiative
cooling and star formation, either without (blue solid line) or with (dotted
green line) additional shear viscosity.  At high redshifts, the signatures of
viscosity are quite similar to the previously considered case where gas
cooling was absent.  However, the formation of a cooling flow for $z < 0.5$
changes the central gas properties dramatically at later times.  Even though
viscous dissipation is acting in the cluster centre, the gas cooling times
become so short there that all the thermal energy gained from internal friction is
radiated away promptly. In fact, the gas starts to cool even more in the
central regions in the viscous case, as can been seen from the entropy
profiles at $z=0$. The gas density is increased in the innermost regions
as well.

Apparently, while the viscous heating in the centre is not sufficiently strong
to raise the temperature significantly, the mild heating does reduce the star
formation rate and therefore the associated non-gravitational heating from
supernovae. The net result is an increase in the X--ray luminosity due to the
sensitive density- and temperature-dependence of the cooling luminosity.
Interestingly, there is still a reduction of the X-ray luminosity in the
outskirts of the cluster in the viscous simulation, because the gas density is
lowered there, but this is just compensated for by a matching increase in the
inner parts.  Another important aspect of the viscous heating process is
that it makes the temperature profile closer to isothermal, suggesting that
the viscosity helps to level the temperature of the cluster on large scales.
It is interesting to note that our results on the profiles resemble the
findings of self-consistent cosmological simulations of cluster formation with
thermal conduction \citep{Jubelgas2004,DolagJubelgas2004}. The
transport coefficients of viscosity and heat conductivity have the same
temperature dependence, and this fact may contribute to the similarity of the
behaviour with respect to the role these transport processes play in
shaping the gas properties.

\begin{figure}
\centerline{
\vbox{
\psfig{file=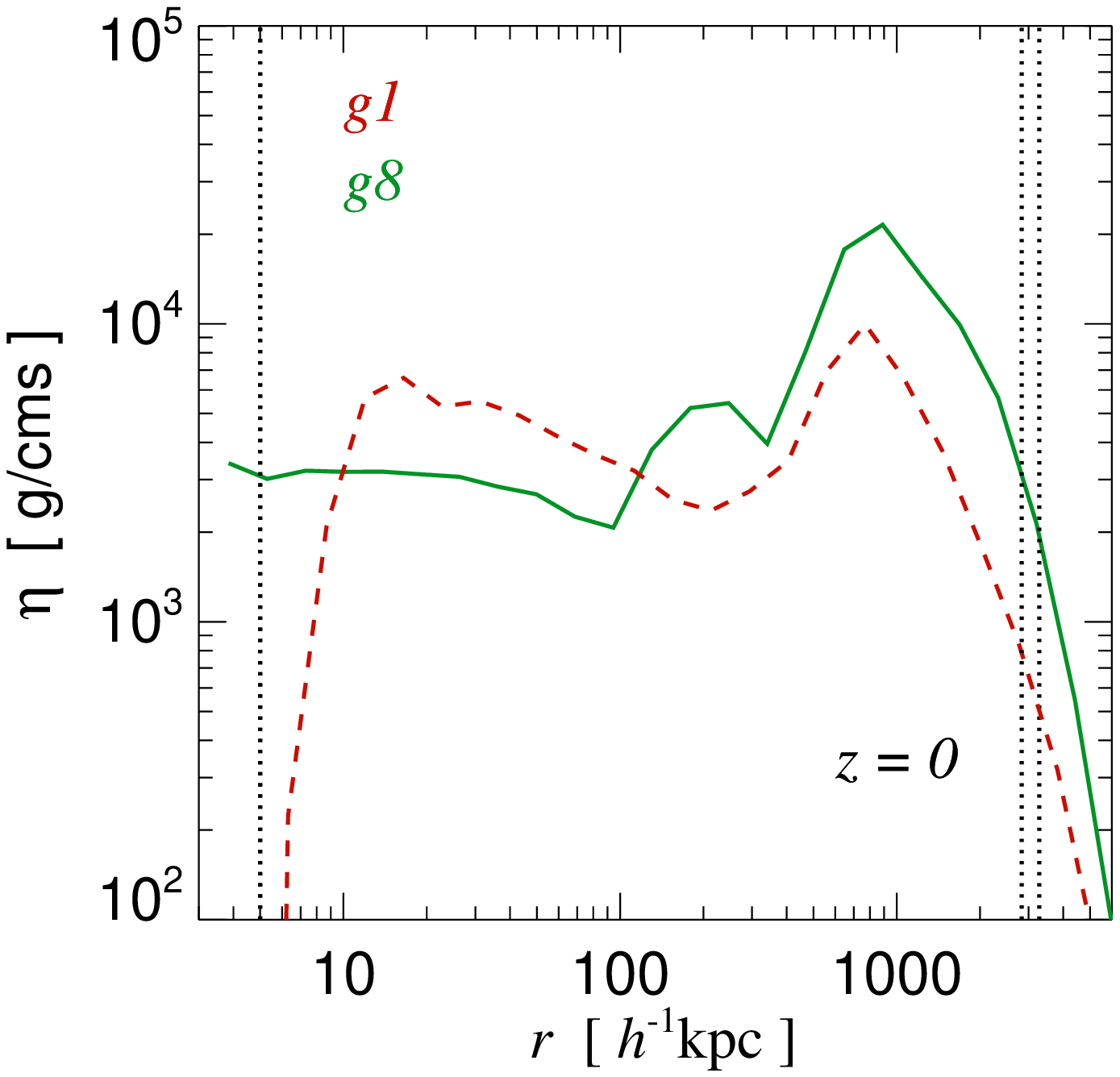,width=8truecm,height=6.7truecm}
\psfig{file=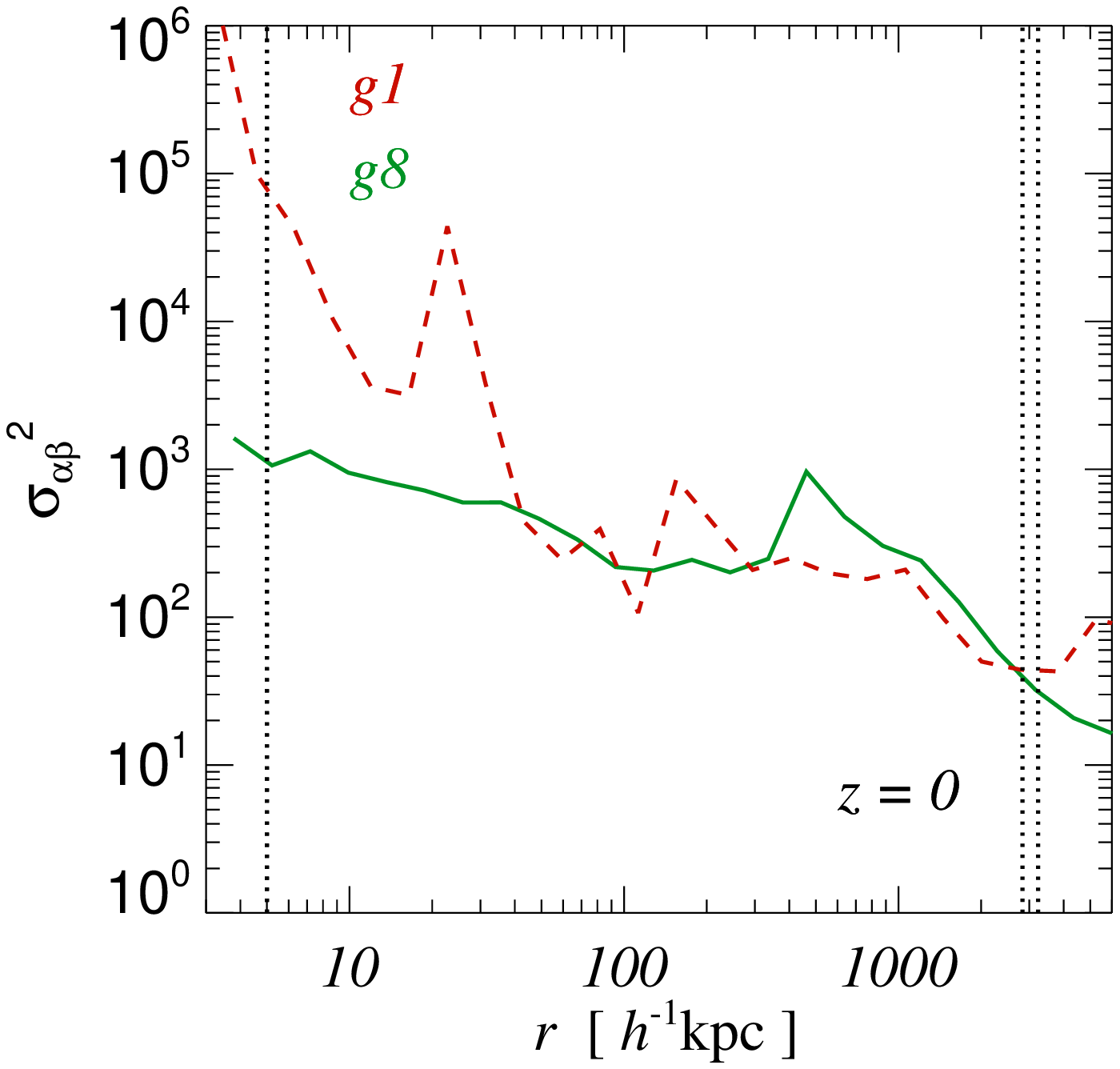,width=8truecm,height=6.7truecm}
\psfig{file=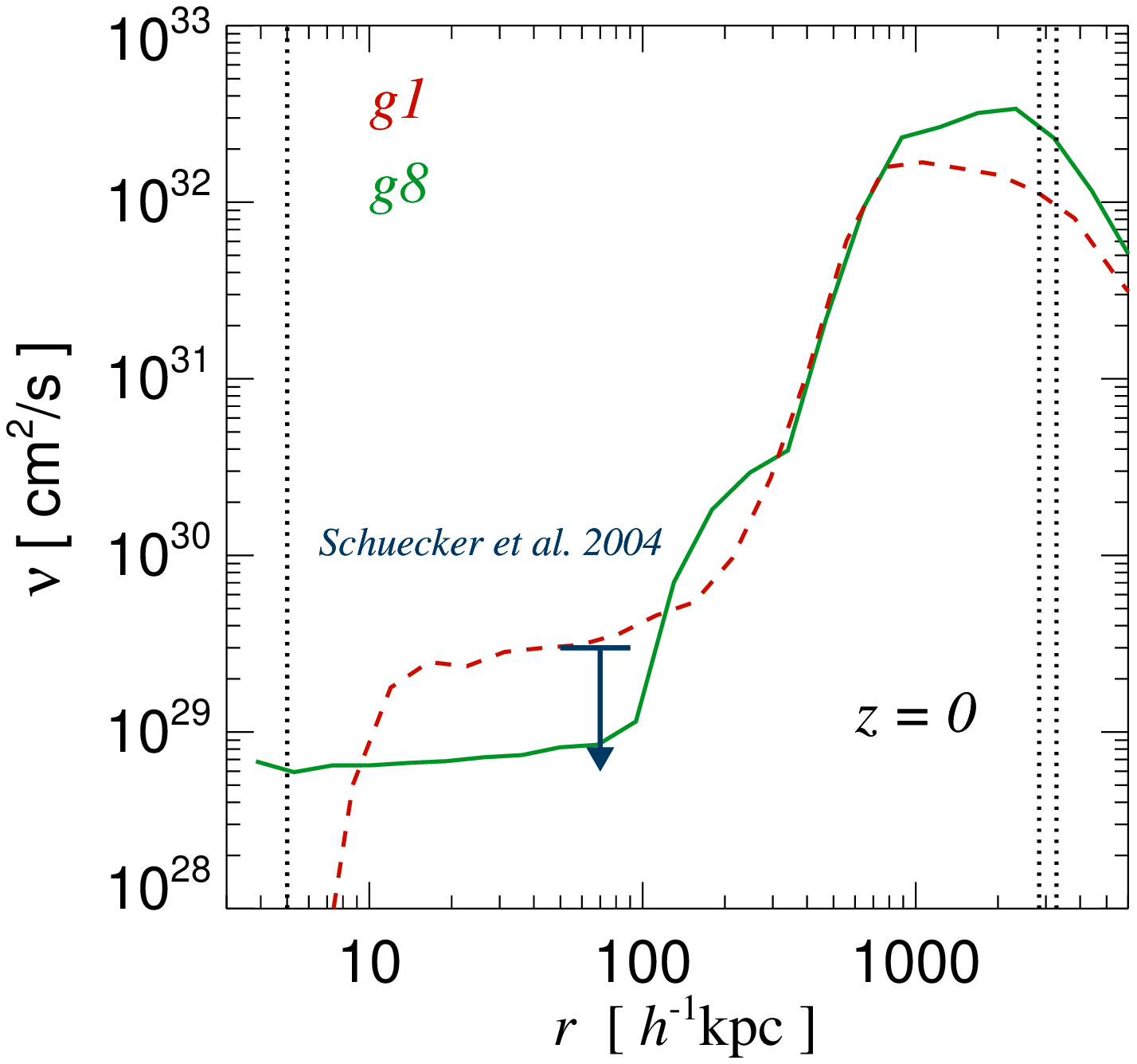,width=8truecm,height=6.7truecm}
}}
\caption{From top to bottom, 
  radial profiles of the mass-weighted shear viscosity coefficient, of the
  rate-of-strain tensor squared, and of the kinematic viscosity coefficient in
  the `g1' and `g8' galaxy cluster simulations, respectively. The red dashed
  line is for the g1 simulation with cooling, star formation and shear
  viscosity, while the continuous green line represents the non-radiative g8
  simulation with the same amount of shear viscosity. The dotted vertical
  lines indicate the gravitational softening and the virial radius at $z=0$,
  respectively. In the bottom panel, the observational upper limit for the
  Coma galaxy cluster \citep{Schuecker2004} at a scale of $100\,{\rm kpc}$ is
  shown with a blue arrow.}
\label{KinemVisc}
\end{figure}

We now turn to a discussion of the radial dependence of viscous dissipation in
clusters of galaxies. As can be seen from equation (\ref{shearvisc}), the
entropy increase due to shear viscosity involves two factors: one is given by
the shear viscosity coefficient, which basically has only a dependence on
temperature to the $5/2$ power, and the other is the ratio of the
rate-of-strain tensor squared to that of the gas density elevated to the
$\gamma$.  Thus, viscous entropy injection will be favoured in regions of high
temperature, low density and strong velocity gradients. In order to
disentangle the relative importance of these different dependencies we show
the radial profiles of the shear viscosity coefficient $\eta$, the
rate-of-strain tensor squared $\sigma_{\alpha \beta}^2$, and of the kinematic
viscosity $\nu$ in Fig.~\ref{KinemVisc}, at $z=0$.

Let us first consider the non--radiative case, which is shown with solid green
lines.  At all radii, the gas density is reduced in the presence of shear
viscosity, and this modification of the density profile influences the rate of
entropy production.  Nevertheless, the contribution of the shear coefficient
is more important in the outer regions, for $r > 100 \,h^{-1}{\rm kpc}$,
having a maximum at $\sim 1000 \,h^{-1}{\rm kpc}$, where also the gas
temperature is highest.  On the other hand, the rate-of-strain tensor is
monotonically increasing from the outskirts towards the centre, and this is
also true for its individual components, indicating that the velocity
gradients are largest in the inner regions.

The dashed red lines show the results when cooling and star formation
are included. The differences that arise in the shear viscosity
coefficient compared with the non-radiative case can be simply
explained by the different temperature profiles: in the outer regions,
for $r > 100 \,h^{-1}{\rm kpc}$, the `g1' cluster run has lower $\eta$
because its temperature there is smaller due to the fact that this
cluster is somewhat less massive than the `g8' galaxy
cluster. Instead, the temperature of the `g1' cluster in the innermost
regions is larger, because the introduction of radiative cooling
increases the central gas temperature, except for the innermost
cooling flow region. Also, $\sigma_{\alpha \beta}^2$ of the `g1'
cluster is higher near the centre due to the motions induced by the
prominent cooling flow that has formed in it.

Finally, we study the kinematic viscosity in the last panel of
Fig.~\ref{KinemVisc}, which is given by the ratio of the shear viscosity
coefficient to the gas density. Both for the `g1' and `g8' cluster
simulations, the kinematic viscosity coefficient is very similar in the outer
regions, and is highest for large radii. In the inner $100 \,h^{-1}{\rm kpc}$,
the kinematic viscosity of the `g1' cluster is higher, due to the efficient
gas cooling. Overplotted with a blue arrow is an upper limit for the kinematic
viscosity on a scale of $100\, {\rm kpc}$ estimated for the Coma galaxy
cluster by \citet{Schuecker2004}. The kinematic viscosity of the `g8' galaxy
cluster, which is slightly more massive then the Coma cluster, is in agreement
with this observational constraint, suggesting that a suppression factor as
large as 0.3 is not ruled out observationally. On the other hand, the
kinematic viscosity of the `g1' cluster, with the same suppression factor, is
only marginally in agreement with the observational upper limit.  However,
this comparison needs to be taken with a grain of salt: The intrinsic amount
of shear viscosity is certainly overestimated in our simulations, because they
suffer from strong central cooling flows which are not observed in real galaxy
clusters.

\subsection{Viscous dissipation during merger events} \label{Merging}

\begin{figure*}
\bc
\centerline{\includegraphics[width=16truecm,height=13truecm]{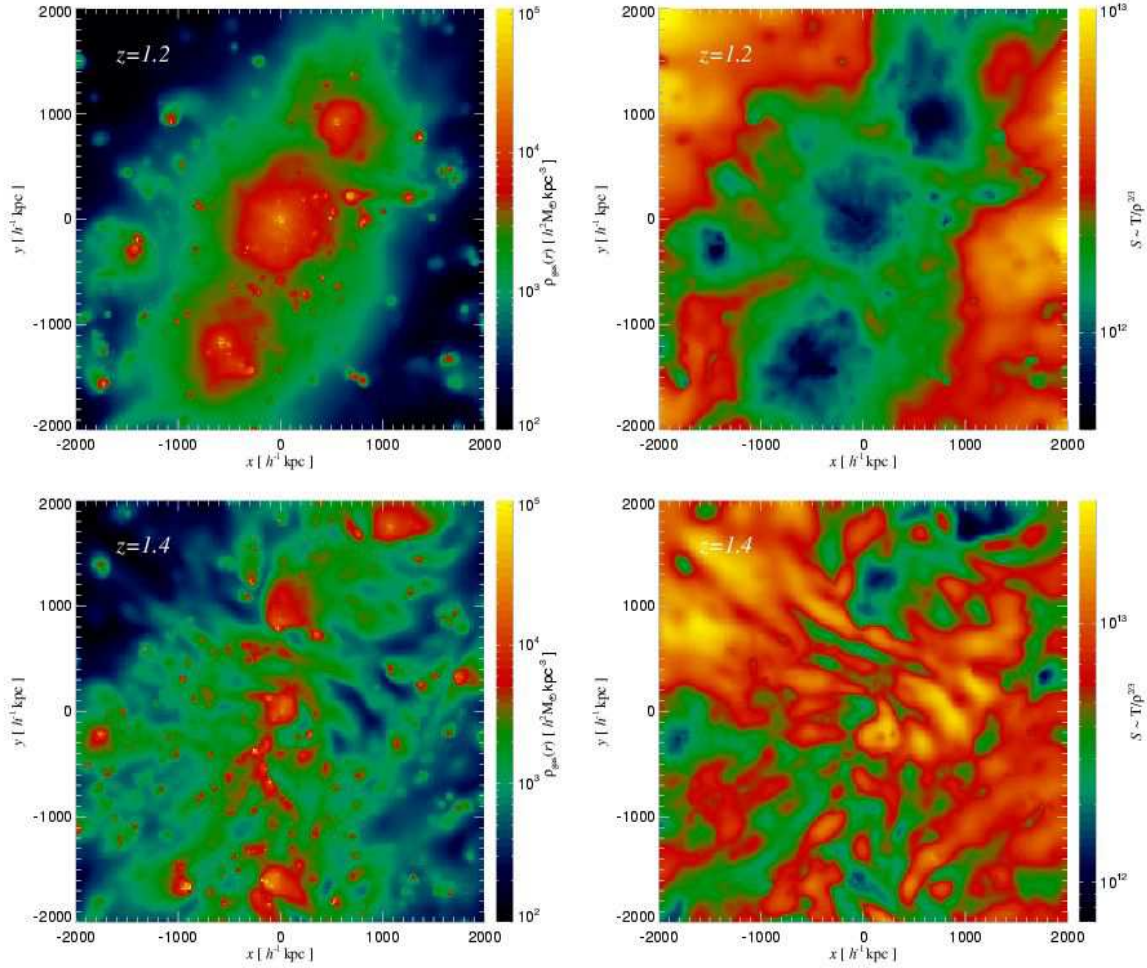}}
\caption{Projected maps of gas density and entropy for the g1 galaxy
  cluster simulation, at an instant just before a merger event. The upper
  panels are for the run with cooling and star formation, while the lower
  panels, at somewhat different redshift, are for the run with additional
  shear viscosity. It can be seen that the entropy of the gas is considerably
  boosted by internal friction processes (note the different scales of the
  entropy maps), and in addition it appears that much of this entropy is
  generated in different regions which have a filamentary kind of structure.}
\label{g1_merger}
\ec
\end{figure*}

The radial profiles of gas temperature and entropy discussed above indicate
that the gas is heated very efficiently in cluster outskirts by viscous
dissipation during accretion and merger events.  In this section, we study
this phenomenon in more detail. In particular, we analyze the spatial
distribution of entropy production just before and during a merger episode. To
this end we compare projected density maps, which give us an indication on the
exact position and extent of the merging clumps, to entropy maps, which tell
us where the heating takes place. In Fig.~\ref{g1_merger}, we show an
example of these maps for the `g1' galaxy cluster. The upper panels
refer to the run with cooling and star formation only, and the lower panels
show the run where shear viscosity was included as well. The panels for the
runs with different physics are not at the same redshift because the shear
forces influence the dynamics of structure formation sufficiently strongly
that timing offsets in the merging histories occur. We  tried however to select
snapshots with similar merging configuration at comparable cosmic epochs so
that the visible differences arise primarily because of the introduction of
the shear viscosity. We note that we analyzed a number of different merger
configurations at multiple redshifts, also considering the non--radiative
simulations. We find that the features visible in Fig.~\ref{g1_merger} are
ubiquitous; viscous dissipation of shear motions not only considerably boosts
the gas entropy, but also generates this entropy in special spatial regions
which have no counterparts in the simulations that only include an artificial
viscosity.  The relevant regions are located perpendicular to the direction of
a halo encounter, and appear as entropy-bright bridges.  These regions of
enhanced entropy, which are never found in the runs without shear viscosity,
are responsible for heating the clusters outskirts, already at early times.

We have also constructed maps of the viscous entropy increase, based on
equation (\ref{shearvisc}). All the filamentary high entropy regions are
corresponding to the ones shown in Fig.~\ref{g1_merger}, demonstrating that
they are caused by $\eta \, \sigma_{\alpha \beta}^2/\rho^{\gamma}$. The
dominant factor appears to be $\eta / \rho^{\gamma}$, while the spatial
distribution of $ \sigma_{\alpha \beta}^2$ is preferentially confined
to the regions characterized by relatively high overdensities. 

\begin{figure*}
\centerline{
\psfig{file=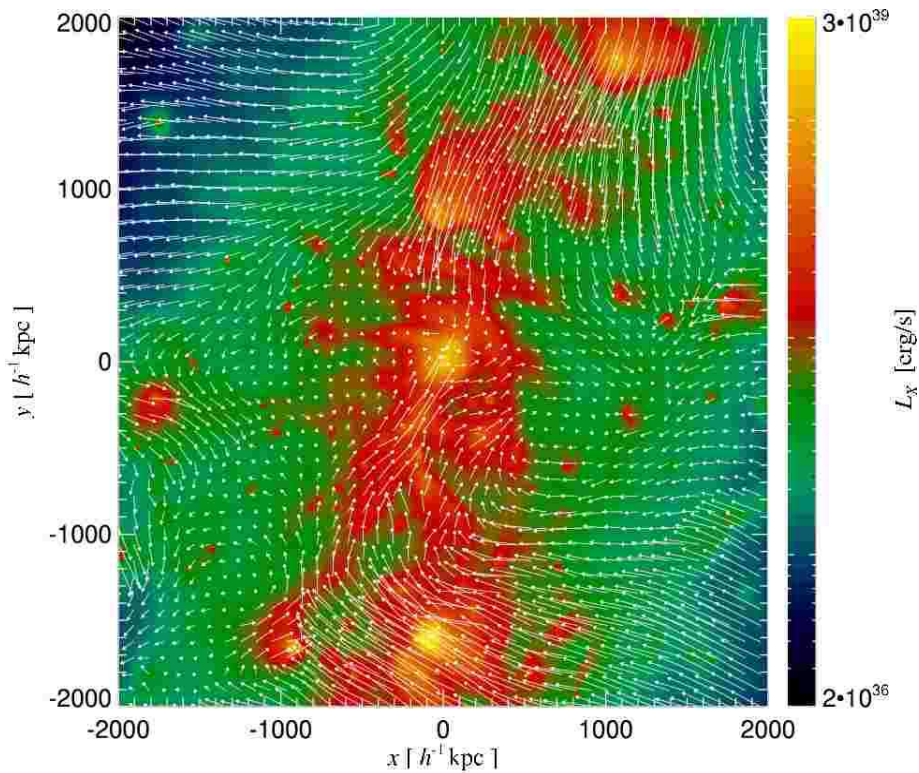,width=12.5truecm,height=11truecm}
}
\caption{Projected X--ray emissivity map of the g1 galaxy cluster,
  simulated with physical viscosity. The map corresponds to the lower panel of
  Fig.~\ref{g1_merger}, but here the velocity field is overplotted with white
  arrows. The flow field suggests that the bright entropy bridges are
  generated in situ due to viscous dissipation.}
\label{vflow_g1}
\end{figure*}

Finally, in order to better understand the gas dynamics in these high
entropy regions, we computed the velocity flow field during the merger
event, and show it overlaid on the X--ray emissivity map in
Fig.~\ref{vflow_g1}.  It can be seen that there are two galaxy
clusters approaching each other, one located in the centre of the
figure and the other one being above it, at $(x,y)\sim
(0,1000)\,h^{-1}{\rm kpc}$. The high entropy bridge is lying between
the merging clusters, roughly perpendicular to their direction of
motion. Considering the velocity flow pattern, it can be noticed that
there are two velocity streams: one starting from the lower right
corner, and the other from the upper right one. The velocity currents
meet in the central region where the merger will occur. Therefore, the
gas is not flowing along the high entropy bridge, but rather
perpendicular to it. This shows that the material of the high entropy
bridge is not funneled towards the centre from the cluster outskirts
but rather that it is heated in situ by significant viscous
dissipation, creating the entropy bridges between the merging
structures.

\section{Discussion and Conclusions} \label{Discussion}

In this study, we discussed a new implementation of viscous fluids in the
parallel TreeSPH code {\small GADGET-2}, in the framework of a self-consistent
entropy and energy conserving formulation of SPH. We presented a discretized
form of the Navier-Stokes and general heat transfer equations, considering
both shear and bulk internal friction forces, subject to a saturation
criterion in order to avoid unphysically large viscous forces. The shear and
bulk viscosity coefficients have either been modeled as being constant, or are
parameterized in the case of shear viscosity with Braginskii's equation,
modified with a suppression factor to describe in a simple fashion a possible
reduction of the effective viscosity due to magnetic fields. Our methodology
for physical viscosity in SPH extends previous works
\citep{Flebbe1994,Schaefer2004} that analyzed viscosity effects in the
context of SPH simulations of planet formation.

We have here applied our new method to simulations of the physics of galaxy
clusters, and in particular to their growth in cosmological simulations of
structure formation.  However, our implementation is general and could, for
example, also be used in studies of accretion disks around black holes,
or for simulations of planet formation.

We have tested our implementation in a number of simple hydrodynamical
problems with known analytic solutions. For example, we simulated
flows between two planes that move with a constant relative velocity,
or that are fixed and embedded in a constant gravitational field. The
stationary solutions we obtained were in good agreement with the
analytic expectations and demonstrated the robustness of our
scheme. We also performed shock tube tests where we investigated the
ability of physical shear and bulk viscosity to capture shocks,
instead of using the artificial viscosity normally invoked in SPH
codes to this end. We found that for fluids with sufficiently high
physical viscosity, shocks can be captured accurately without an
artificial viscosity. However, since in practical applications the
physical viscosity is often comparatively low, an additional
artificial viscosity is still indicated in most cases. In particular,
for our applications in cosmological simulations where we invoked
Braginskii parameterization for the shear viscosity, we are dealing
with an effective viscosity which is a strong function of gas
temperature. If an artificial viscosity was omitted, simulations would
suffer from particle interpenetration at low temperatures.

We have applied our new numerical scheme to study the influence of viscosity
on clusters of galaxies. We first considered the physics of hot, buoyant
bubbles in the ICM, which are injected by an AGN.  In a previous study
\citep{Sijacki2006} we have already studied this type of feedback in some
detail, but we were here interested in the specific question to what extent
the introduction of a certain amount of shear viscosity changes the
interaction of the bubbles with the surrounding ICM. We found that the AGN
bubbles can still heat the intracluster gas efficiently in the viscous case,
but depending on the strength of the viscosity, the properties of the ICM in
the inner $\sim 150 \,h^{-1}{\rm kpc}$ are altered. The viscous dissipation of
sound waves generated by the bubbles does not result in a significant
non-local heating of the ICM. To confirm this claim, we estimated the
energy budget and the spatial extent of the sound waves for different
amounts of viscosity. We found, in agreement with a previous estimate of
\cite{Churazov2002}, that the total energy in sound waves is only a small
fraction of the initial bubble energy. However, a caveat lies in
the fact that the initial stages of bubble generation by the AGN-jets are not
modeled in our simulation. It is conceivable that these early stages provide
stronger sound waves with a potentially bigger impact on the ICM.

We also analyzed bubble morphologies, dynamics and survival times as a
function of increasing shear viscosity. Similar to previous analytic and
numerical works \citep{Kaiser2005,Reynolds2005} we find that an increasing gas
viscosity stabilizes bubbles against Kelvin-Helmholtz and Rayleigh-Taylor
instabilities, delaying their shredding. Thus, bubbles can rise further away
from the cluster centre for the same initial specific entropy content.
Because we simulated a long time span, we could follow many bubble duty
cycles. This allowed us to conclude that the observation of multiple bubble
episodes, as is the case in the Perseus cluster \citep[e.g.][]{Fabian2006},
can be used to infer a minimum value for the ICM gas viscosity, otherwise the
bubbles could not survive for such a long time. This constraint is however
weakened by the possibility that magnetic fields or relativistic particle
populations change the dynamics of the bubbles.

Finally, we addressed the role of gas viscosity in the context of
cosmological simulations of galaxy cluster formation.  Using a set of
non--radiative cluster simulations, we showed that already a modest
level of shear viscosity (with a suppression factor of $0.3$) has a
profound effect on galaxy clusters.  The gas density distribution is
significantly changed compared to the case where only an artificial
viscosity is included, with substructures loosing their baryons more
easily, leaving behind a large number of prominent gaseous tails. Only
recently XMM-Newton and Chandra observations \citep{Wang2004,
Sun2005,Sun2006} have started discovering long (from $60\,{\rm kpc}$
to $88 \,{\rm kpc}$) and narrow ($< 16 \,{\rm kpc}$) X-ray tails
behind late type galaxies in hot clusters.  Further observational
studies could put constraints on the amount of gas viscosity by
analyzing the X-ray features of the tails. In principle, this could be
directly compared with our numerical simulations when synthetic X-ray
emissivity maps are constructed.  Another interesting imprint of
internal friction processes was found in merger episodes of
clusters. The entropy of the gas is not just boosted everywhere by a
fixed amount due to the viscous dissipation, but instead the entropy
increase occurs in filament-like structures which have no
corresponding counterparts in simulations where only artificial
viscosity is included.

When the physics of radiative gas cooling is included as well, the
effects of internal friction remain similar in the cluster periphery,
while in the innermost regions, the radiative cooling timescale
becomes so short that the whole energy liberated by internal friction
is radiated away.  Therefore, in cosmological simulations of cluster
formation with radiative cooling, gas viscosity cannot prevent the
formation of a central cooling flow. On the other hand, the general
tendency of gas viscosity to flatten the temperature profile, and to
boost the gas entropy in the outskirts, brings the simulations of
galaxy clusters closer to observational results. At least one other
physical ingredient is needed to simultaneously solve the
over--cooling problem while keeping the benefits of viscous
effects. AGN feedback appears to be a promising candidate in this
respect, but it remains a complex task to construct a fully
self-consistent simulation model that includes all these processes
accurately, and with a minimum of assumptions.

\section*{Acknowledgements}
We are grateful to Simon White and Eugene Churazov for very
constructive discussions and comments on the manuscript. We thank
Klaus Dolag for providing initial conditions of galaxy clusters. DS
acknowledges the PhD fellowship of the International Max Planck
Research School in Astrophysics, and the support received from a Marie
Curie Host Fellowship for Early Stage Research Training.

\bibliographystyle{mnras}

\bibliography{paper}

\end{document}